\documentclass[mathleft
]{an}
\usepackage{graphicx}
\usepackage{times}
\newcommand{\logg}{$\log g$}
\newcommand{\teff}{$T_{\rm eff}$}
\newcommand{\solarmass}{${\rm M}_\odot$}
\newcommand{\lowm}{${\rm M}<3$\,${\rm M}_\odot$}
\newcommand{\highm}{${\rm M}>3$\,${\rm M}_\odot$}
\newcommand{\highmx}[2]{${\rm M}>{#1}$\,${\rm M}_\odot$}

\begin{document}

\Pagespan{1}{}
\Yearpublication{2007}%
\Yearsubmission{2007}%
\Month{11}%
\Volume{999}%
\Issue{88}%

\title{Evolution of Magnetic Fields in Stars Across the Upper Main Sequence: \\
II.\ Observed distribution of the magnetic field geometry}

\author{
S. Hubrig\inst{1}\fnmsep\thanks{Corresponding author: shubrig@eso.org\newline}
\and  P. North\inst{2}
\and  M. Sch\"oller\inst{1}
}

\titlerunning{Evolution of Magnetic Fields}
\authorrunning{S. Hubrig et al.}
\institute{
European Southern Observatory, Casilla 19001, Santiago 19, Chile
\and 
Laboratoire d'Astrophysique, Ecole Polytechnique F\'ed\'erale
de Lausanne (EPFL), Observatoire,
CH-1290~Sauverny, Switzerland
}

\received{30 Ene 2007}
\accepted{11 Mar 2007}
\publonline{later}

\abstract{%
We re-discuss the evolutionary state of upper main sequence magnetic stars
using a sample of Ap and Bp stars with accurate Hipparcos parallaxes
and definitely determined longitudinal magnetic fields. 
We confirm our previous results obtained from the study of Ap and Bp stars
with accurate measurements of the mean magnetic field modulus and mean
quadratic magnetic fields that magnetic stars
of mass \lowm{} are concentrated towards the centre of the main-sequence band.
In contrast, stars with masses \highm{} seem to be concentrated closer 
to the ZAMS. The study of a few known members of nearby open clusters with accurate 
Hipparcos parallaxes confirms these conclusions.
Stronger magnetic fields tend to be found in hotter, younger
and more massive stars, as well as in stars with shorter rotation periods. 
The longest rotation periods are 
found only in stars which spent already more than 40\% of their main sequence 
life, in the mass domain between 
1.8 and 3\,\solarmass{} and with \logg{} values  ranging from 3.80 to 4.13.
No evidence is found for any loss of angular momentum during the main-sequence life.
The magnetic flux remains constant over the stellar life time on the main sequence.
An excess of stars with large obliquities $\beta$ is detected in both higher and 
lower mass stars.
It is quite possible that the angle $\beta$ becomes close to 0$^\circ$ in slower 
rotating stars of mass \highm{} too, 
analog to the behaviour of angles $\beta$ in slowly rotating stars of \lowm{}.
The obliquity angle distribution as inferred from the distribution of $r$-values
appears random at the time magnetic stars become observable on the H-R diagram. After quite 
a short time spent on the  main sequence, the obliquity angle $\beta$ tends to 
reach  values close to either 90$^\circ$ or 0$^\circ$ for \lowm{}.
The evolution of the obliquity angle $\beta$ seems to be somewhat different for 
low and high mass stars. While we find a strong hint 
for an increase of $\beta$ with the elapsed time on the main sequence for stars 
with \highm{},
no similar trend is found for stars with \lowm{}. However,
the predominance of high values of $\beta$ at advanced ages in these stars is notable. 
As the physics governing the processes taking place in magnetised atmospheres remains 
poorly understood, magnetic field properties have to be 
considered in the framework of dynamo or fossil field theories.
}

\keywords{stars:chemically peculiar -- stars:evolution -- stars:magnetic fields 
-- stars:fundamental parameters -- stars:rotation -- stars:statistics}

\maketitle

\section{Introduction}

After presenting in the first part of this paper the results of a magnetic field study of more than one hundred  
chemically peculiar (CP) A and B stars with previously unknown or only poorly determined magnetic fields 
using FORS\,1 at the 
Very Large Telescope (Hubrig et al.\ \cite{hu06}), we examine in this second part of our paper the evolution 
of the magnetic field across the upper main sequence (MS). The emphasis is put on the study of the 
evolutionary aspect of the magnetic field geometry in stars with accurately determined positions in the 
H-R diagram and well-studied periodic magnetic variations.
Knowledge of the evolution of the magnetic field and its geometry, especially of the 
distribution of the obliquity angle $\beta$ (orientation of the magnetic axis with respect 
to the rotation axis) is
essential to understand the physical
processes taking place in these stars and the origin of their magnetic fields (fossil versus 
contemporary dynamo). 
Early studies of the evolution of some characteristics of magnetic Ap and Bp stars were based on smaller 
stellar samples (e.g., North \cite{no85}, Hubrig, North \& Mathys \cite{hu00a}, Hubrig, North \& Szeifert \cite{hu05}), or on 
stellar samples which included 
CP stars with not well-defined magnetic field strengths or rotation periods 
(e.g., R\"udiger \& Scholz \cite{ru88}, 
P\"ohnl, Paunzen \& Maitzen \cite{po05}, Kochukhov \& Bagnulo \cite{ko06}).

The determination of rotation periods of Ap and Bp stars is in some cases not an easy task, as some stars
are extremely slowly rotating with periods up to 70~years and longer. 
The fact that magnetic stars are slow 
rotators suggests that some mechanism of magnetic braking has been working in a certain 
phase of the stellar evolution. However, the comparison of 
rotation periods of Ap stars 
of different ages did not present any evidence supporting the 
hypothesis that Ap stars suffer considerable magnetic braking during their MS life 
(e.g., North \cite{no85}, North \cite{nort98}, Hubrig, North \& Mathys \cite{hu00a}).
Also the search for progenitors of slowly rotating magnetic Ap stars among normal A stars
showed no statistically significant difference between the $v$\,sin\,$i$ 
distributions of young and old A-type stars (Hubrig, North \& Medici \cite{hu00b}).

The periodic magnetic variations of Ap and Bp stars are generally described
by the oblique rotator model (Stibbs \cite{stibbs}) with a magnetic field 
having an axis of symmetry at an angle 
to the rotation axis, usually called $\beta$. If the magnetic field is excited by 
a dynamo mechanism, the magnetic field can be either symmetric or antisymmetric in regard to 
the equatorial plane (Krause \& Oetken \cite{krause}).  In case the stars have 
acquired their magnetic field at the time of their formation or early in their evolution, the 
currently observed magnetic field is called a fossil field. Randomness in the $\beta$- 
distribution is usually regarded as an argument in favour of the fossil theory, as 
the star-to-star variations in obliquity of magnetic field axes can plausibly be interpreted as 
reflecting differences in the intrinsic magnetic conditions at different formation sites.
Clearly, the decision between the fossil and dynamo explanations for the magnetic field origin
can be made only through careful consolidation of magneto-hydrodynamic theory and observations.
Our finding that no strong magnetic fields are found in A stars close to the ZAMS with masses 
\lowm{}, supports those theories that locate a dynamo in the convective core since there should be 
enough time in the star's evolution to transport the core dynamo field to the stellar surface.
As hotter, more massive magnetic Bp stars are mainly found to be located on the ZAMS or very 
close to it (Hubrig, North \& Szeifert \cite{hu05}) it is quite possible that Bp stars acquired their magnetic field 
already at the time of their formation.
The previous weakness of the fossil-field theory has been the absence of magnetic field configurations
stable enough to survive in a star over its lifetime.
Braithwaite \& Spruit (\cite{braith}) recently carried out numerical simulations to show that 
stable magnetic field configurations can develop through evolution from arbitrary, unstable initial fields.
The possibility of a magnetic field resulting from a magneto-rotational 
instability in the radiative envelope has been discussed by
R\"udiger, Arlt \& Hollerbach (\cite{ruediger}) and Arlt (\cite{arlt}).
Over the years, a lot of work has been done to show that
differential rotation, advection by internal circulation,
stellar wind, magnetic torques, and rotation about a non-principal axis can all produce
 changes in the $\beta$ angle of an oblique rotator. Although the time scales over 
which these (and other) processes cause $\beta$ to change remain fairly uncertain, 
it could be possible that the observed magnetic field distributions reflect not only 
how the magnetic field was produced but its subsequent evolution in response to the
operation of various mechanisms.

\section{Basic data}

\begin{table*}
\caption{
Fundamental parameters of stars with \highm{}.
}
\label{tab:1}
\begin{center}
\begin{tabular}{rrrrrrrrrr}
\hline
\multicolumn{1}{c}{HD} & \multicolumn{1}{c}{M$_V$} & \multicolumn{1}{c}{$M$/M$_\odot$} & \multicolumn{1}{c}{log $T_{\rm eff}$} & \multicolumn{1}{c}{log $L$/L$_\odot$} & \multicolumn{1}{c}{\logg} & \multicolumn{1}{c}{$R$/R$_\odot$} & \multicolumn{1}{c}{d[pc]} & \multicolumn{1}{c}{$\sigma(\pi)/\pi$} & \multicolumn{1}{c}{$f$} \\
\hline
5737 & -2.27 & 4.976$\pm$0.335 & 4.121$\pm$0.013 & 3.068$\pm$0.156 & 3.50$\pm$0.14 & 6.54$\pm$1.24 & 206. & 0.173 & 1.15 \\
12767 & -0.54 & 3.643$\pm$0.152 & 4.111$\pm$0.013 & 2.369$\pm$0.085 & 4.03$\pm$0.09 & 3.06$\pm$0.35 & 111. & 0.086 & 0.57 \\
18296 & -0.44 & 3.273$\pm$0.143 & 4.036$\pm$0.013 & 2.271$\pm$0.101 & 3.78$\pm$0.10 & 3.87$\pm$0.51 & 119. & 0.107 & 0.84 \\
19832 & 0.34 & 3.142$\pm$0.144 & 4.095$\pm$0.013 & 2.008$\pm$0.096 & 4.26$\pm$0.10 & 2.17$\pm$0.27 & 114. & 0.100 & 0.13 \\
21699 & -1.06 & 4.314$\pm$0.235 & 4.159$\pm$0.013 & 2.660$\pm$0.118 & 4.00$\pm$0.11 & 3.44$\pm$0.51 & 180. & 0.127 & 0.61 \\
22470 & -0.67 & 3.736$\pm$0.179 & 4.115$\pm$0.013 & 2.424$\pm$0.111 & 4.00$\pm$0.11 & 3.20$\pm$0.45 & 145. & 0.119 & 0.61 \\
24155 & 0.30 & 3.353$\pm$0.176 & 4.132$\pm$0.013 & 2.059$\pm$0.112 & 4.38$\pm$0.11 & 1.95$\pm$0.28 & 136. & 0.121 & 0.01 \\
25823 & -0.48 & 3.615$\pm$0.234 & 4.112$\pm$0.026 & 2.343$\pm$0.119 & 4.05$\pm$0.15 & 2.95$\pm$0.54 & 152. & 0.129 & 0.54 \\
28843 & 0.05 & 3.555$\pm$0.173 & 4.143$\pm$0.013 & 2.179$\pm$0.103 & 4.33$\pm$0.10 & 2.13$\pm$0.28 & 131. & 0.109 & 0.01 \\
34452 & -0.32 & 3.754$\pm$0.212 & 4.158$\pm$0.026 & 2.270$\pm$0.093 & 4.33$\pm$0.14 & 2.20$\pm$0.35 & 137. & 0.096 & 0.01 \\
40312 & -0.99 & 3.385$\pm$0.088 & 4.006$\pm$0.026 & 2.370$\pm$0.055 & 3.57$\pm$0.12 & 4.98$\pm$0.67 & 53. & 0.043 & 1.03 \\
49333 & -0.56 & 4.289$\pm$0.262 & 4.182$\pm$0.013 & 2.538$\pm$0.134 & 4.21$\pm$0.12 & 2.68$\pm$0.44 & 205. & 0.148 & 0.22 \\
73340 & -0.11 & 3.667$\pm$0.130 & 4.145$\pm$0.013 & 2.253$\pm$0.068 & 4.28$\pm$0.08 & 2.30$\pm$0.23 & 143. & 0.063 & 0.07 \\
79158 & -0.94 & 3.820$\pm$0.217 & 4.097$\pm$0.013 & 2.521$\pm$0.131 & 3.84$\pm$0.12 & 3.89$\pm$0.63 & 176. & 0.144 & 0.79 \\
92664 & -0.32 & 3.863$\pm$0.147 & 4.154$\pm$0.013 & 2.357$\pm$0.075 & 4.24$\pm$0.09 & 2.47$\pm$0.26 & 143. & 0.073 & 0.16 \\
103192 & -0.56 & 3.387$\pm$0.139 & 4.044$\pm$0.013 & 2.335$\pm$0.094 & 3.76$\pm$0.09 & 4.00$\pm$0.50 & 112. & 0.099 & 0.85 \\
112413 & 0.25 & 3.021$\pm$0.096 & 4.060$\pm$0.015 & 2.031$\pm$0.050 & 4.08$\pm$0.08 & 2.63$\pm$0.24 & 34. & 0.035 & 0.50 \\
122532 & -0.16 & 3.271$\pm$0.166 & 4.071$\pm$0.013 & 2.206$\pm$0.117 & 3.98$\pm$0.11 & 3.05$\pm$0.45 & 169. & 0.127 & 0.64 \\
124224 & 0.47 & 3.036$\pm$0.115 & 4.084$\pm$0.013 & 1.954$\pm$0.074 & 4.25$\pm$0.09 & 2.15$\pm$0.23 & 80. & 0.072 & 0.15 \\
125823 & -1.19 & 5.651$\pm$0.255 & 4.248$\pm$0.013 & 3.032$\pm$0.094 & 4.10$\pm$0.10 & 3.50$\pm$0.43 & 128. & 0.098 & 0.43 \\
133652 & 0.74 & 3.050$\pm$0.132 & 4.113$\pm$0.013 & 1.863$\pm$0.089 & 4.46$\pm$0.09 & 1.69$\pm$0.20 & 96. & 0.092 & 0.01 \\
133880 & 0.20 & 3.143$\pm$0.151 & 4.079$\pm$0.013 & 2.075$\pm$0.102 & 4.13$\pm$0.10 & 2.54$\pm$0.33 & 127. & 0.108 & 0.41 \\
137509 & -0.29 & 3.367$\pm$0.203 & 4.076$\pm$0.022 & 2.268$\pm$0.138 & 3.95$\pm$0.14 & 3.21$\pm$0.60 & 249. & 0.152 & 0.67 \\
142301 & -0.28 & 4.243$\pm$0.289 & 4.193$\pm$0.013 & 2.470$\pm$0.151 & 4.32$\pm$0.14 & 2.36$\pm$0.43 & 140. & 0.168 & 0.01 \\
142990 & -0.80 & 4.902$\pm$0.260 & 4.217$\pm$0.013 & 2.765$\pm$0.114 & 4.18$\pm$0.11 & 2.97$\pm$0.43 & 150. & 0.123 & 0.28 \\
144334 & -0.31 & 4.085$\pm$0.256 & 4.167$\pm$0.024 & 2.465$\pm$0.118 & 4.20$\pm$0.15 & 2.65$\pm$0.46 & 149. & 0.128 & 0.25 \\
147010 & 0.59 & 3.137$\pm$0.180 & 4.117$\pm$0.013 & 1.931$\pm$0.125 & 4.42$\pm$0.12 & 1.80$\pm$0.28 & 143. & 0.136 & 0.01 \\
151965 & -0.10 & 3.736$\pm$0.245 & 4.154$\pm$0.013 & 2.272$\pm$0.145 & 4.31$\pm$0.13 & 2.25$\pm$0.40 & 181. & 0.161 & 0.01 \\
168733 & -1.16 & 4.015$\pm$0.230 & 4.108$\pm$0.020 & 2.614$\pm$0.131 & 3.81$\pm$0.14 & 4.12$\pm$0.73 & 190. & 0.144 & 0.81 \\
173650 & -0.48 & 3.095$\pm$0.192 & 4.000$\pm$0.013 & 2.191$\pm$0.144 & 3.69$\pm$0.13 & 4.17$\pm$0.73 & 215. & 0.159 & 0.90 \\
175362 & -0.38 & 3.986$\pm$0.267 & 4.164$\pm$0.030 & 2.404$\pm$0.114 & 4.24$\pm$0.17 & 2.50$\pm$0.47 & 130. & 0.123 & 0.16 \\
196178 & -0.13 & 3.542$\pm$0.162 & 4.126$\pm$0.013 & 2.227$\pm$0.096 & 4.22$\pm$0.10 & 2.43$\pm$0.30 & 147. & 0.100 & 0.21 \\
223640 & 0.17 & 3.192$\pm$0.149 & 4.089$\pm$0.013 & 2.081$\pm$0.097 & 4.17$\pm$0.10 & 2.44$\pm$0.31 & 98. & 0.102 & 0.33 \\
\hline
\end{tabular}
\end{center}
\end{table*}

\begin{table*}
\caption{
Fundamental parameters of stars with \lowm{}.
}
\label{tab:2}
\begin{center}
\begin{tabular}{rrrrrrrrrr}
\hline
\multicolumn{1}{c}{HD} & \multicolumn{1}{c}{M$_V$} & \multicolumn{1}{c}{$M$/M$_\odot$} & \multicolumn{1}{c}{log $T_{\rm eff}$} & \multicolumn{1}{c}{log $L$/L$_\odot$} & \multicolumn{1}{c}{\logg} & \multicolumn{1}{c}{$R$/R$_\odot$} & \multicolumn{1}{c}{d[pc]} & \multicolumn{1}{c}{$\sigma(\pi)/\pi$} & \multicolumn{1}{c}{$f$} \\
\hline
2453 & 0.91 & 2.289$\pm$0.112 & 3.940$\pm$0.015 & 1.600$\pm$0.113 & 3.91$\pm$0.11 & 2.78$\pm$0.41 & 152. & 0.121 & 0.74 \\
3980 & 1.63 & 1.975$\pm$0.065 & 3.916$\pm$0.016 & 1.296$\pm$0.052 & 4.05$\pm$0.09 & 2.19$\pm$0.21 & 65. & 0.038 & 0.57 \\
4778 & 1.23 & 2.242$\pm$0.085 & 3.972$\pm$0.014 & 1.486$\pm$0.072 & 4.14$\pm$0.09 & 2.10$\pm$0.22 & 91. & 0.069 & 0.42 \\
8441 & 0.25 & 2.621$\pm$0.167 & 3.964$\pm$0.014 & 1.868$\pm$0.147 & 3.80$\pm$0.13 & 3.39$\pm$0.62 & 204. & 0.163 & 0.83 \\
9996 & 0.83 & 2.433$\pm$0.126 & 3.987$\pm$0.013 & 1.660$\pm$0.110 & 4.06$\pm$0.11 & 2.40$\pm$0.34 & 139. & 0.119 & 0.54 \\
10783 & 0.06 & 2.832$\pm$0.155 & 4.006$\pm$0.013 & 1.988$\pm$0.126 & 3.88$\pm$0.12 & 3.21$\pm$0.50 & 186. & 0.138 & 0.76 \\
12288 & 0.61 & 2.486$\pm$0.161 & 3.972$\pm$0.014 & 1.741$\pm$0.150 & 3.93$\pm$0.13 & 2.82$\pm$0.52 & 231. & 0.166 & 0.70 \\
12447 & 1.05 & 2.423$\pm$0.075 & 4.008$\pm$0.013 & 1.587$\pm$0.056 & 4.22$\pm$0.08 & 2.00$\pm$0.17 & 43. & 0.045 & 0.25 \\
14437 & 0.48 & 2.802$\pm$0.206 & 4.034$\pm$0.013 & 1.906$\pm$0.164 & 4.07$\pm$0.15 & 2.57$\pm$0.51 & 198. & 0.184 & 0.52 \\
15144 & 1.99 & 1.878$\pm$0.071 & 3.922$\pm$0.016 & 1.155$\pm$0.067 & 4.20$\pm$0.09 & 1.81$\pm$0.19 & 66. & 0.062 & 0.33 \\
24712 & 2.54 & 1.604$\pm$0.058 & 3.859$\pm$0.018 & 0.904$\pm$0.054 & 4.13$\pm$0.09 & 1.81$\pm$0.19 & 49. & 0.041 & 0.50 \\
32633 & 0.92 & 2.533$\pm$0.174 & 4.021$\pm$0.025 & 1.667$\pm$0.133 & 4.21$\pm$0.16 & 2.07$\pm$0.39 & 157. & 0.146 & 0.26 \\
49976 & 1.28 & 2.195$\pm$0.090 & 3.962$\pm$0.014 & 1.458$\pm$0.081 & 4.12$\pm$0.09 & 2.13$\pm$0.24 & 101. & 0.081 & 0.46 \\
54118 & 0.36 & 2.744$\pm$0.094 & 4.022$\pm$0.017 & 1.883$\pm$0.053 & 4.03$\pm$0.09 & 2.64$\pm$0.26 & 87. & 0.040 & 0.57 \\
62140 & 1.92 & 1.822$\pm$0.070 & 3.882$\pm$0.017 & 1.163$\pm$0.065 & 4.01$\pm$0.10 & 2.20$\pm$0.24 & 81. & 0.059 & 0.64 \\
65339 & 1.26 & 2.076$\pm$0.070 & 3.916$\pm$0.016 & 1.411$\pm$0.077 & 3.96$\pm$0.09 & 2.49$\pm$0.29 & 98. & 0.076 & 0.69 \\
71866 & 0.90 & 2.287$\pm$0.117 & 3.937$\pm$0.015 & 1.600$\pm$0.118 & 3.90$\pm$0.11 & 2.82$\pm$0.43 & 147. & 0.128 & 0.75 \\
72968 & 1.11 & 2.251$\pm$0.086 & 3.960$\pm$0.014 & 1.526$\pm$0.073 & 4.06$\pm$0.09 & 2.33$\pm$0.25 & 82. & 0.070 & 0.55 \\
74521 & 0.08 & 2.984$\pm$0.129 & 4.033$\pm$0.013 & 2.066$\pm$0.100 & 3.93$\pm$0.10 & 3.10$\pm$0.40 & 125. & 0.105 & 0.70 \\
81009 & 1.56 & 1.961$\pm$0.089 & 3.894$\pm$0.017 & 1.313$\pm$0.104 & 3.95$\pm$0.11 & 2.47$\pm$0.35 & 139. & 0.111 & 0.71 \\
83368 & 1.98 & 1.794$\pm$0.068 & 3.876$\pm$0.017 & 1.137$\pm$0.062 & 4.01$\pm$0.10 & 2.19$\pm$0.23 & 72. & 0.055 & 0.64 \\
90044 & 0.76 & 2.514$\pm$0.099 & 4.002$\pm$0.013 & 1.707$\pm$0.078 & 4.09$\pm$0.09 & 2.36$\pm$0.26 & 108. & 0.078 & 0.50 \\
90569 & 0.65 & 2.519$\pm$0.114 & 3.993$\pm$0.013 & 1.733$\pm$0.094 & 4.03$\pm$0.10 & 2.54$\pm$0.32 & 118. & 0.098 & 0.58 \\
94660 & 0.16 & 2.945$\pm$0.121 & 4.032$\pm$0.013 & 2.037$\pm$0.095 & 3.95$\pm$0.09 & 3.02$\pm$0.38 & 152. & 0.099 & 0.67 \\
96707 & 0.90 & 2.224$\pm$0.068 & 3.891$\pm$0.017 & 1.575$\pm$0.069 & 3.73$\pm$0.09 & 3.38$\pm$0.38 & 109. & 0.065 & 0.90 \\
98088 & 0.87 & 2.264$\pm$0.093 & 3.905$\pm$0.016 & 1.604$\pm$0.094 & 3.76$\pm$0.10 & 3.27$\pm$0.43 & 129. & 0.098 & 0.87 \\
108945 & 0.57 & 2.431$\pm$0.085 & 3.944$\pm$0.015 & 1.725$\pm$0.079 & 3.82$\pm$0.09 & 3.16$\pm$0.36 & 95. & 0.079 & 0.80 \\
111133 & 0.29 & 2.651$\pm$0.156 & 3.978$\pm$0.014 & 1.878$\pm$0.136 & 3.85$\pm$0.12 & 3.22$\pm$0.54 & 161. & 0.149 & 0.78 \\
112185 & -0.21 & 2.832$\pm$0.054 & 3.953$\pm$0.015 & 2.038$\pm$0.042 & 3.62$\pm$0.07 & 4.33$\pm$0.36 & 25. & 0.015 & 0.96 \\
116114 & 1.49 & 1.953$\pm$0.098 & 3.865$\pm$0.018 & 1.327$\pm$0.116 & 3.81$\pm$0.12 & 2.86$\pm$0.45 & 140. & 0.125 & 0.83 \\
116458 & -0.12 & 2.947$\pm$0.103 & 4.012$\pm$0.013 & 2.071$\pm$0.080 & 3.83$\pm$0.08 & 3.44$\pm$0.38 & 142. & 0.080 & 0.79 \\
118022 & 1.15 & 2.234$\pm$0.073 & 3.960$\pm$0.014 & 1.509$\pm$0.056 & 4.07$\pm$0.08 & 2.29$\pm$0.21 & 56. & 0.045 & 0.53 \\
119213 & 1.55 & 2.055$\pm$0.073 & 3.944$\pm$0.015 & 1.335$\pm$0.063 & 4.14$\pm$0.09 & 2.02$\pm$0.20 & 88. & 0.056 & 0.42 \\
119419 & 1.16 & 2.297$\pm$0.116 & 3.979$\pm$0.023 & 1.533$\pm$0.084 & 4.13$\pm$0.13 & 2.15$\pm$0.31 & 113. & 0.084 & 0.43 \\
125248 & 1.35 & 2.203$\pm$0.091 & 3.972$\pm$0.014 & 1.442$\pm$0.082 & 4.18$\pm$0.09 & 2.00$\pm$0.23 & 90. & 0.082 & 0.35 \\
126515 & 1.21 & 2.246$\pm$0.139 & 3.970$\pm$0.014 & 1.497$\pm$0.135 & 4.13$\pm$0.13 & 2.15$\pm$0.36 & 141. & 0.149 & 0.44 \\
128898 & 2.10 & 1.804$\pm$0.055 & 3.903$\pm$0.016 & 1.100$\pm$0.041 & 4.16$\pm$0.08 & 1.86$\pm$0.16 & 16. & 0.010 & 0.41 \\
133029 & 0.54 & 2.710$\pm$0.139 & 4.033$\pm$0.024 & 1.816$\pm$0.082 & 4.14$\pm$0.13 & 2.32$\pm$0.34 & 146. & 0.082 & 0.40 \\
137909 & 1.17 & 2.091$\pm$0.044 & 3.872$\pm$0.018 & 1.460$\pm$0.045 & 3.74$\pm$0.08 & 3.24$\pm$0.31 & 35. & 0.024 & 0.89 \\
137949 & 1.94 & 1.781$\pm$0.061 & 3.847$\pm$0.019 & 1.150$\pm$0.077 & 3.88$\pm$0.10 & 2.54$\pm$0.31 & 89. & 0.076 & 0.80 \\
140160 & 1.11 & 2.191$\pm$0.063 & 3.934$\pm$0.015 & 1.509$\pm$0.065 & 3.96$\pm$0.08 & 2.57$\pm$0.26 & 70. & 0.059 & 0.69 \\
143473 & 1.00 & 2.505$\pm$0.156 & 4.021$\pm$0.025 & 1.637$\pm$0.114 & 4.24$\pm$0.15 & 2.00$\pm$0.35 & 124. & 0.123 & 0.20 \\
148112 & 0.19 & 2.639$\pm$0.082 & 3.958$\pm$0.014 & 1.888$\pm$0.071 & 3.76$\pm$0.08 & 3.57$\pm$0.38 & 72. & 0.068 & 0.86 \\
148199 & 0.43 & 2.883$\pm$0.169 & 4.046$\pm$0.013 & 1.947$\pm$0.128 & 4.09$\pm$0.12 & 2.54$\pm$0.40 & 151. & 0.140 & 0.49 \\
148330 & 0.50 & 2.504$\pm$0.069 & 3.965$\pm$0.014 & 1.766$\pm$0.062 & 3.88$\pm$0.08 & 2.99$\pm$0.29 & 112. & 0.055 & 0.75 \\
152107 & 1.17 & 2.187$\pm$0.067 & 3.944$\pm$0.015 & 1.488$\pm$0.047 & 4.02$\pm$0.08 & 2.40$\pm$0.21 & 54. & 0.028 & 0.61 \\
153882 & 0.14 & 2.688$\pm$0.133 & 3.966$\pm$0.014 & 1.921$\pm$0.114 & 3.76$\pm$0.11 & 3.57$\pm$0.52 & 169. & 0.123 & 0.85 \\
164258 & 0.61 & 2.383$\pm$0.109 & 3.908$\pm$0.016 & 1.707$\pm$0.105 & 3.69$\pm$0.11 & 3.64$\pm$0.52 & 121. & 0.111 & 0.93 \\
170000 & 0.26 & 2.991$\pm$0.099 & 4.058$\pm$0.015 & 2.009$\pm$0.054 & 4.09$\pm$0.08 & 2.58$\pm$0.24 & 89. & 0.043 & 0.48 \\
170397 & 1.32 & 2.281$\pm$0.087 & 3.993$\pm$0.013 & 1.471$\pm$0.075 & 4.25$\pm$0.09 & 1.88$\pm$0.20 & 87. & 0.073 & 0.19 \\
187474 & 0.35 & 2.691$\pm$0.102 & 4.004$\pm$0.013 & 1.872$\pm$0.087 & 3.96$\pm$0.09 & 2.83$\pm$0.33 & 104. & 0.089 & 0.66 \\
188041 & 0.99 & 2.215$\pm$0.077 & 3.909$\pm$0.016 & 1.557$\pm$0.079 & 3.81$\pm$0.09 & 3.05$\pm$0.36 & 85. & 0.079 & 0.83 \\
192678 & 0.39 & 2.424$\pm$0.115 & 3.959$\pm$0.014 & 1.701$\pm$0.110 & 3.91$\pm$0.11 & 2.86$\pm$0.41 & 230. & 0.118 & 0.73 \\
196502 & -0.35 & 2.871$\pm$0.091 & 3.939$\pm$0.015 & 2.099$\pm$0.072 & 3.50$\pm$0.09 & 4.96$\pm$0.54 & 128. & 0.069 & 1.19 \\
201601 & 1.97 & 1.807$\pm$0.061 & 3.882$\pm$0.017 & 1.144$\pm$0.049 & 4.03$\pm$0.09 & 2.15$\pm$0.21 & 35. & 0.032 & 0.62 \\
208217 & 1.57 & 1.967$\pm$0.104 & 3.899$\pm$0.016 & 1.313$\pm$0.121 & 3.97$\pm$0.12 & 2.41$\pm$0.38 & 146. & 0.132 & 0.68 \\
220825 & 1.45 & 2.123$\pm$0.067 & 3.958$\pm$0.014 & 1.383$\pm$0.053 & 4.16$\pm$0.08 & 2.00$\pm$0.18 & 50. & 0.039 & 0.39 \\
\hline
\end{tabular}
\end{center}
\end{table*}

To study the evolution of the magnetic field distribution in Ap and Bp stars we decided 
to compile Hipparcos parallaxes and photometric data exclusively for magnetic Ap and Bp stars
with well studied periodic variations of their mean longitudinal magnetic fields. 
The mean longitudinal magnetic field is an average over the visible stellar
hemisphere of the component of the magnetic vector along the line of sight, and 
is derived from measurements of wavelength shifts of spectral lines between
right and left circular polarisation. The individual published measurements of the mean
longitudinal magnetic field $\left< B_l \right>$ have recently been put together in the catalog of stellar magnetic 
rotational phase curves $\left< B_l \right>(\phi)$ by Bychkov, Bychkova \& Madej (\cite{bych05}).
The phase $\phi$ is determined by

\begin{equation}
\phi = 2 \pi \left( \frac{T_i-T_0}{P} \right),
\end{equation}

\noindent
with $T_i$ the time of measurement, $T_0$ the zero 
epoch, i.e.\ the time corresponding to the phase $\phi=0$, and $P$ the rotation period.
Out of 127 Ap and Bp stars presented in this catalog, only for 89 stars with well defined 
magnetic curves the parallax error is
less than 20\% and Str\"omgren and/or Geneva photometry is available. One more star, HD\,81009, with an
accurate Hipparcos parallax and a well-defined magnetic rotational phase curve studied 
by Landstreet \& Mathys (\cite {lama00}) has also been included in our sample. 

Our results of the study of the evolutionary state of upper main sequence magnetic stars
using a smaller sample of Ap and Bp stars with accurate Hipparcos parallaxes
and definitely determined longitudinal magnetic fields
indicated a notable difference between the distributions of Ap and Bp stars in the H-R diagram.
In contrast to magnetic stars of mass
\lowm{} which have mostly been found around the centre of the main-sequence band, 
stars with masses \highm{} seemed to be concentrated closer to the ZAMS, and the stronger 
magnetic fields tend to be found in hotter, younger (in terms of the elapsed fraction of main-sequence
life) and more massive stars (Hubrig, North \& Szeifert \cite{hu05}).  
In view of these apparent differences in the evolutionary state between magnetic stars in 
different mass domains, 
we decided to group our sample in two sub-samples with one containing lower mass stars with \lowm{}
(57 stars in all) and another one with \highm{} (33 stars).
The effective temperatures of the stars of both sub-samples have been determined
from photometric data in the Geneva system through the calibration of K\"unzli et al.\ (\cite{k97})
corrected according to the prescription of North (\cite{nor98}), or,
when no Geneva photometry was available, in the
Str\"omgren system, applying the calibration of Moon \& Dworetsky (\cite{mn85}),
revised by Napiwotski et al.\ (\cite{nap93}).
The luminosities have been obtained by taking into account the bolometric
corrections measured by Lanz (\cite{La84}).
In case of binary systems, a correction for the duplicity has been applied to their
magnitudes (i.e., $0.75$\,mag for SB2 and $0.3$\,mag for SB1 systems).
The Lutz--Kelker correction (Lutz \& Kelker \cite{lk73}) has not been used, as it still 
remains a matter of debate (e.g., Arenou \& Luri \cite{arenou}).
The stellar masses are obtained by interpolation in the evolutionary tracks of Schaller et al. 
(1992) for a solar metallicity $Z = 0.018$, as explained in North, Jaschek \& Egret (\cite{nje97}).
The radii have been computed from the luminosity and effective temperature.
Finally, the surface gravity was obtained from mass and radius through its fundamental definition. 
The errors on \teff{} and $L$ determination were linearised and carefully propagated to errors on mass, 
radius, and \logg{}, taking into account the slope of the evolutionary tracks.
More details on the determination of the fundamental parameters can be found
in our previous paper (Hubrig, North \& Mathys \cite{hu00a}). 

The basic data for both sub-samples are presented in Tables~\ref{tab:1} and \ref{tab:2}.
The columns are, in order,
the HD number of the star, the absolute visual magnitude, the mass (in solar masses),
the logarithm of the effective temperature, the logarithm of the luminosity 
(relative to the Sun), the logarithm of the stellar gravity, the radius (in solar 
radii), the distance $d$, the relative uncertainty of the parallax, and the fraction 
$f$ of the main-sequence life completed by the star, interpolated from theoretical 
evolutionary tracks (Schaller et al.\ \cite{schaller}). 
For a number of stars located below the ZAMS the masses have been calculated 
by assuming their position on the ZAMS. Since ages are ill-defined for stars close to
the ZAMS, in many applications we used gravity as a proxy for the relative age 
which  has also the advantage of being a more directly measured quantity.
For stars located close to the terminal-age main sequence (TAMS) we estimated an ambiguity of 
the order of 5\% on the interpolated 
mass as there exists the possibility that these stars are making a loop back to the blue.

\section{Analysis and results}

\subsection{The distribution of the magnetic Ap and Bp stars in the H-R diagram}

\begin{figure}
\centering
{\includegraphics[width=0.45\textwidth]{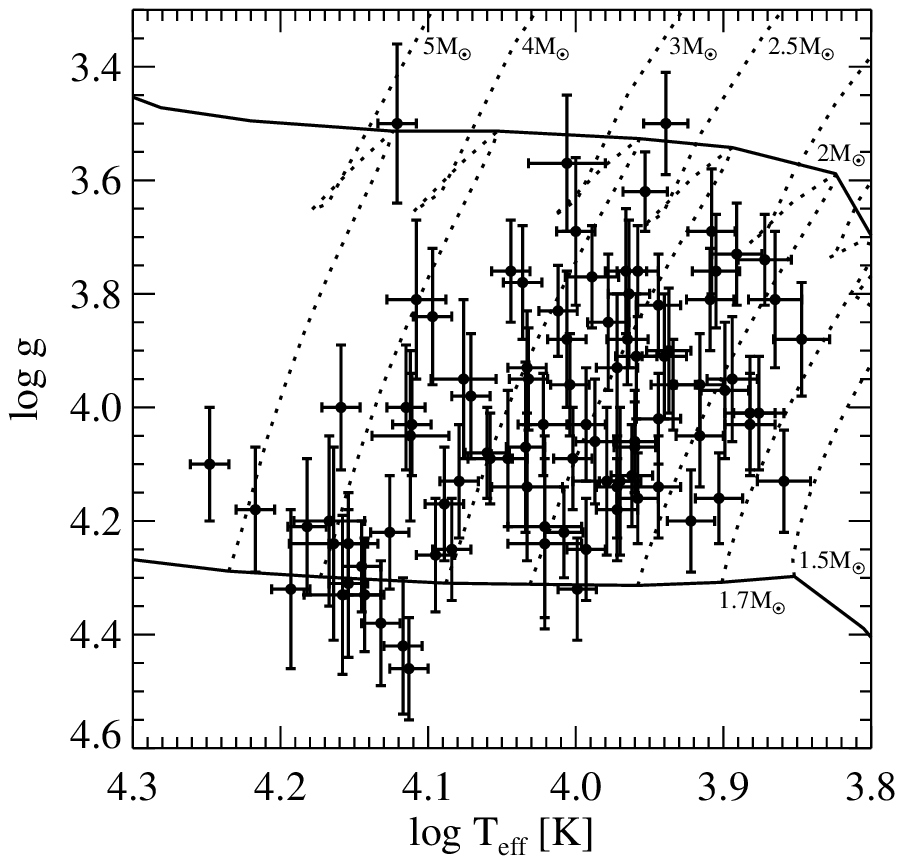}}
\caption{H-R digram for Ap and Bp stars with accurate Hipparcos parallaxes and well studied 
periodic variations of their mean longitudinal magnetic fields. The vertical and horizontal 
bars indicate the accuracy of the determination of effective temperature and gravity.}
\label{fig00a}
\end{figure}

\begin{figure}
\centering
{\includegraphics[width=0.45\textwidth]{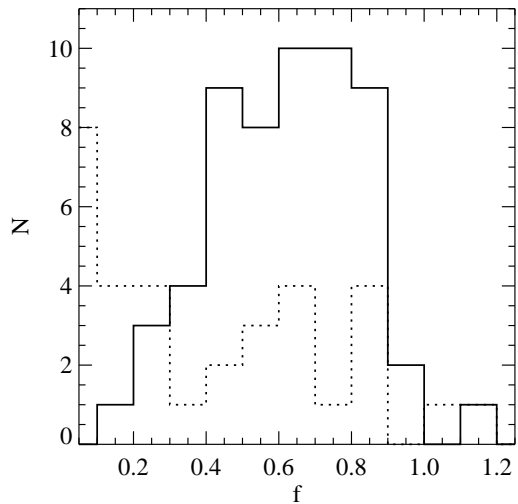}}
\caption{Distribution of relative ages for stars with masses \highm{} (dotted
line) and \lowm{} (solid line).
}
\label{fig00b}
\end{figure}

\begin{figure}
\centering
{\includegraphics[width=0.40\textwidth]{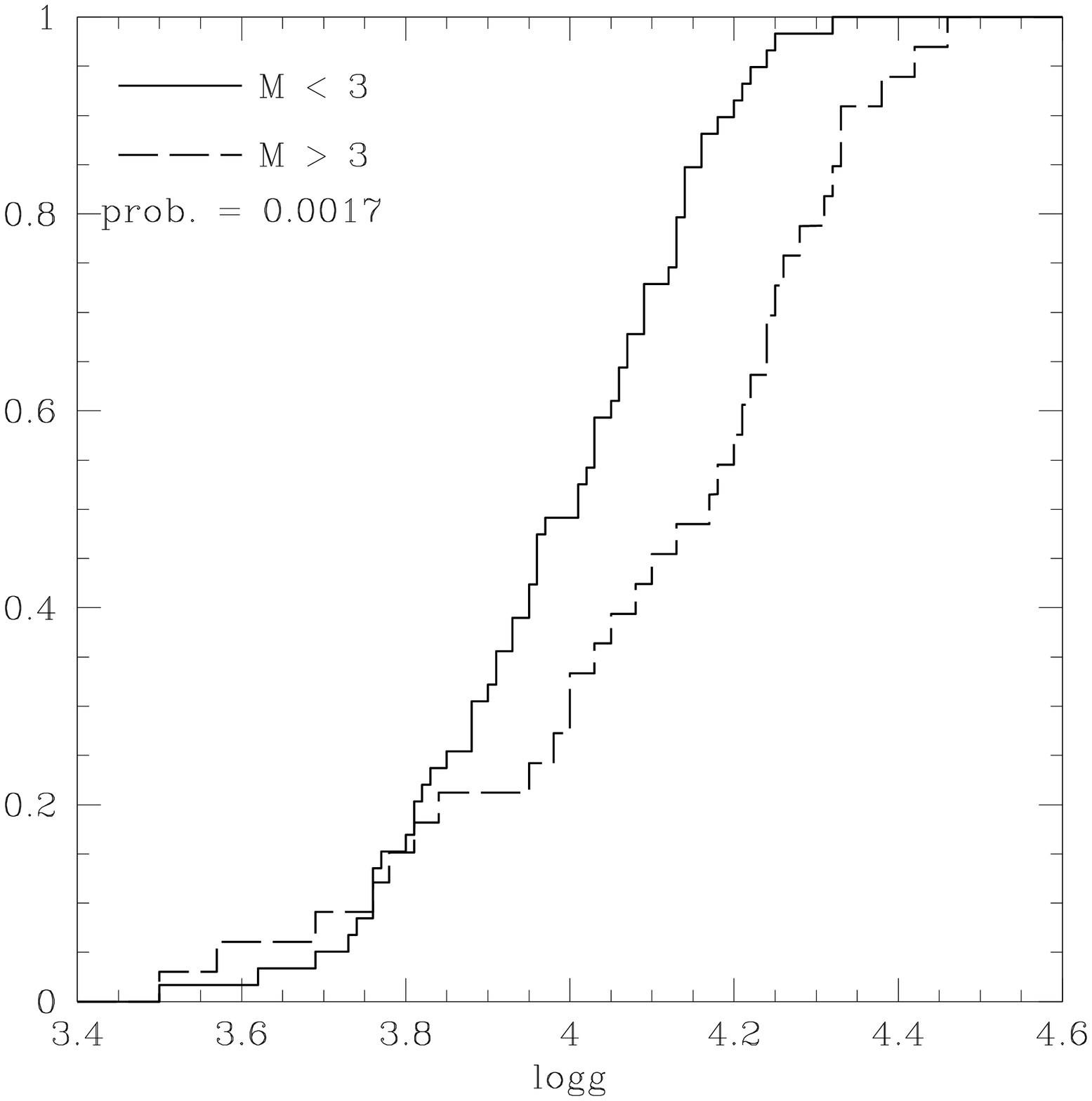}}
\caption{Cumulative distribution of $\log_g$-values for magnetic stars with masses \highm{} (dashed
line) and \lowm{} (solid line).
}
\label{fig00c}
\end{figure}

The position of the studied magnetic Ap and Bp stars in the H-R diagram is shown in 
Fig.~\ref{fig00a}.
From this figure it is quite clear 
that the magnetic stars with \lowm{} show a concentration towards the centre 
of the main-sequence band and are rarely found close to either 
ZAMS or TAMS, in full agreement  with our previous results 
(Hubrig, North \& Mathys \cite{hu00a}, Hubrig, North \& Szeifert \cite{hu05}).
By contrast, the massive Bp stars are much younger, with a
distribution crowded towards the ZAMS. We note that a number of stars in the mass domain 
between 3 and 4\,\solarmass{}  are distributed  well across the main-sequence band.
The distribution of relative ages for each group is presented in Fig.~\ref{fig00b}.
and shows a broad maximum indicating that the majority of magnetic stars with \lowm{}
have completed 40 to 90\% of their main-sequence life. The situation is completely different 
for stars with \highm{}: the distribution peaks at the ZAMS age with two secondary lower 
peaks at the relative ages around 60\% and 80\%. However, given the small number of stars in 
this sample, the  conclusion about secondary peaks
can be considered as only marginally significant and should be confirmed by larger statistics.

Out of our sample of 57 stars with \lowm{}, only three Ap stars, HD\,12447, HD\,32633, and 
HD\,143473, have relative ages between 20 and 26\%. 
In the following we show the difficulty of determining their fundamental parameters.
HD\,12447 with a relative age of 25\% is a visual double star with $\Delta{}m = 1$. 
The (B-V)$_{\rm Tycho}$ for this star was transformed to the (B-V)$_{\rm Johnson}$
through the relation (B-V)$_{\rm Johnson}$ = 0.850 (B-V)$_{\rm Tycho}$ given in
ESA (\cite{esa97}), then to (B-V)$_{\rm Geneva}$ through the relation given by
Meylan \& Hauck (\cite{mey81}).
The secondary has an index typical of an A5V
star, the average colours of which (Hauck \cite{hau94}) were
subtracted from the colours of the binary to obtain the colours of the
primary.
Then \teff{} was obtained from X and Y and corrected according to
the scheme proposed by North (\cite{nor98}).
The \teff{} determination for HD\,32633 with a relative age of 26\% is very uncertain,
ranging from our photometric estimate of 12,819\,K to 10,250\,K using the (B2-G) index (Sokolov \cite{so98}). 
We assumed \teff{} = 10,500\,K, which corresponds to the spectral type B9 given by Borra \& Landstreet 
(\cite{bo80}). 
The reddening A$_{\rm V}$ = 0.16 was adopted following Lucke's maps, instead
of the initial A$_{\rm V}$ = 0.411 obtained from the X,Y Geneva parameters. 
The star HD\,143473 with a relative age 
of 20\% has already been discussed in our previous paper (Hubrig, North \& Mathys \cite{hu00a}). 
The reddening obtained from the Geneva X, Y parameters appears much too high
(A$_{\rm V}$ = 0.95) in view of the modest distance,
and the effective temperature for this star was therefore determined from spectroscopy rather from than 
from photometry. This star is also an extreme photometric Ap star with
a photometric peculiarity parameter
$\Delta(V1-G) = 0.048$ (Hauck \& North \cite{ha81}).

In Fig.~\ref{fig00c} we show the cumulative distribution of \logg{} for both sub-samples. A 
Kolmogorov-Smirnov test implies that the distribution of \logg{} values for more massive stars differs 
from the distribution of stars with \lowm{} at a significance level of 99.8\%.

\subsection{The evolution of the magnetic field strength across the main sequence}

\begin{table}
\caption{
Model parameters for stars with \highm{}.
}
\label{tab:3}
\begin{center}
\begin{tabular}{rrrrrrr}
\hline
\multicolumn{1}{c}{\raisebox{2mm}{\rule{0mm}{2mm}}HD} &
\multicolumn{1}{c}{$\overline{\left< B_l \right>}$} &
\multicolumn{1}{c}{$P_{\rm rot}$} &
\multicolumn{1}{c}{$v$\,sin\,$i$} &
\multicolumn{1}{c}{$r$} & \multicolumn{1}{c}{$i$} & \multicolumn{1}{c}{$\beta$} \\
\hline
5737 & 324 & 21.65 & 20 & -0.583 & 88.51 & 5.64 \\
12767 & 242 & 1.89 & 45 & -0.743 & 33.32 & 84.46 \\
18296 & 440 & 2.88 & 25 & -0.618 & 21.57 & 84.67 \\
19832 & 315 & 0.73 & 110 & -0.710 & 47.00 & 79.69 \\
21699 & 828 & 2.48 & 35 & -0.803 & 29.91 & 86.40 \\
22470 & 733 & 0.68 & 60 & -0.819 & 14.59 & 88.52 \\
24155 & 1034 & 2.53 & 35 & -0.280 & 63.82 & 41.15 \\
25823 & 668 & 4.66 & 25 & -0.214 & 51.30 & 51.05 \\
28843 & 344 & 1.37 & 100 & -0.670 & 89.92 & 0.39 \\
34452 & 743 & 2.47 & 46 & -0.371 & 86.83 & 6.88 \\
40312 & 223 & 3.62 & 49 & -0.619 & 44.74 & 76.87 \\
49333 & 618 & 2.18 & 65 & -0.780 & 85.34 & 33.40 \\
73340 & 1644 & 2.67 & -- & 0.307 & \multicolumn{1}{c}{--} & \multicolumn{1}{c}{--} \\
79158 & 672 & 3,83 & 60 & -0.910 & 66.13 & 83.92 \\
92664 & 803 & 1.67 & 66 & 0.163 & 61.87 & 21.04 \\
103192 & 204 & 2.34 & 72 & 0.301 & 56.35 & 19.68 \\
112413 & 1349 & 5.47 & 10 & -0.670 & 24.27 & 84.91 \\
122532 & 665 & 3.68 & 35 & -0.919 & 56.57 & 86.34 \\
124224 & 572 & 0.52 & 119 & -0.510 & 34.67 & 77.35 \\
125823 & 469 & 8.82 & 15 & -0.766 & 48.33 & 81.53 \\
133652 & 1116 & 2.30 & 69 & -0.703 & 84.36 & 29.52 \\
133880 & 2415 & 0.88 & -- & -0.848 & \multicolumn{1}{c}{--} & \multicolumn{1}{c}{--} \\
137509 & 1021 & 4.49 & 28 & -0.689 & 50.72 & 77.31 \\
142301 & 2104 & 1.46 & 58 & -0.430 & 45.16 & 68.16 \\
142990 & 1304 & 0.98 & 125 & -0.438 & 54.60 & 61.19 \\
144334 & 783 & 1.49 & 44 & -0.194 & 29.27 & 69.28 \\
147010 & 4032 & 3.92 & 22 & 0.460 & 71.24 & 7.16 \\
151965 & 2603 & 1.61 & 105 & 0.216 & 88.85 & 0.74 \\
168733 & 815 & 14.78 & 12 & 0.306 & 58.29 & 18.17 \\
173650 & 326 & 9.98 & 15 & -0.466 & 45.19 & 69.86 \\
175362 & 3570 & 3.67 & 15 & -0.614 & 25.80 & 83.41 \\
196178 & 973 & 1.92 & 50 & 0.089 & 51.33 & 33.80 \\
223640 & 643 & 3.74 & 30 & -0.064 & 65.34 & 27.56 \\
\hline
\end{tabular}
\end{center}
\end{table}

\begin{table}
\caption{
Model parameters for stars with \lowm{}.
}
\label{tab:4}
\begin{center}
\begin{tabular}{rrrrrrr}
\hline
\multicolumn{1}{c}{\raisebox{2mm}{\rule{0mm}{2mm}}HD} &
\multicolumn{1}{c}{$\overline{\left< B_l \right>}$} &
\multicolumn{1}{c}{$P_{\rm rot}$} &
\multicolumn{1}{c}{$v$\,sin\,$i$} &
\multicolumn{1}{c}{$r$} & \multicolumn{1}{c}{$i$} & \multicolumn{1}{c}{$\beta$} \\
\hline
2453 & 588 & 521 & 0 & 0.450 & 62.00 & 11.00 \\
3980 & 1202 & 3.95 & 15 & -0.921 & 32.32 & 88.51 \\
4778 & 1026 & 2.56 & 33 & -0.764 & 52.66 & 80.05 \\
8441 & 284 & 1.81 & 4 & -0.357 & 2.42 & 88.85 \\
9996 & 833 & 7692 & 0 & \multicolumn{1}{c}{--} & \multicolumn{1}{c}{--} & \multicolumn{1}{c}{--} \\
10783 & 1269 & 4.15 & 17 & 0.115 & 25.74 & 58.72 \\
11503 & 545 & 1.61 & 13 & -0.479 & 6.47 & 87.71 \\
12288 & 1643 & 34.90 & 0 & 0.171 & 62.00 & 22.00 \\
12447 & 365 & 1.49 & 56 & -0.810 & 55.54 & 81.30 \\
14437 & 1824 & 26.8 & 0 & 0.360 & 56.00 & 19.00 \\
15144 & 803 & 3.0 & 18 & 0.727 & 36.13 & 12.22 \\
24712 & 803 & 12.46 & 5.6 & 0.199 & 49.63 & 29.60 \\
32633 & 2761 & 6.43 & 20 & -0.499 & 88.03 & 5.86 \\
49976 & 1489 & 2.98 & 23 & -0.825 & 39.49 & 85.48 \\
54118 & 1033 & 5.63 & -- & -0.924 & \multicolumn{1}{c}{--} & \multicolumn{1}{c}{--} \\
62140 & 1336 & 4.28 & 18 & -0.934 & 43.79 & 88.13 \\
65339 & 3113 & 8.03 & 12.5 & -0.920 & 50.00 & 86.00 \\
71866 & 1678 & 6.80 & 8.9 & -0.802 & 25.10 & 87.05 \\
72968 & 480 & 4.66 & 11 & -0.733 & 25.77 & 85.75 \\
74521 & 813 & 7.77 & 19 & 0.660 & 70.25 & 4.21 \\
81009 & 1431 & 33.96 & 5.0 & 0.640 & 48.00 & 11.00 \\
83368 & 577 & 2.85 & 32.6 & -0.893 & 56.97 & 85.03 \\
90044 & 739 & 4.38 & 15 & -0.863 & 33.38 & 87.23 \\
90569 & 193 & 1.45 & 18 & -0.399 & 11.72 & 84.91 \\
94660 & 2354 & 2700 & 7.0 & 0.900 & 47.00 & 5.00 \\
96707 & 1072 & 2.49 & 39 & -0.602 & 34.60 & 80.28 \\
98088 & 802 & 5.82 & 13.9 & -0.743 & 29.27 & 85.28 \\
108945 & 537 & 1.92 & 64 & -0.408 & 50.22 & 63.21 \\
111133 & 807 & 16.31 & 4.6 & 0.270 & 27.42 & 47.93 \\
112185 & 93 & 5.09 & 34 & -0.509 & 52.17 & 67.26 \\
116114 & 1923 & 27.6 & 1.0 & 0.920 & 56.00 & 2.00 \\
116458 & 1925 & 148.40 & 0 & 0.680 & 52.00 & 10.00 \\
118022 & 808 & 3.72 & 7.0 & 0.163 & 12.99 & 72.23 \\
119213 & 1220 & 2.45 & 25 & -0.096 & 36.82 & 58.31 \\
119419 & 1770 & 2.60 & 34.8 & -0.595 & 56.27 & 69.18 \\
125248 & 1505 & 9.30 & 6.7 & -0.931 & 38.00 & 88.40 \\
126515 & 1723 & 129.95 & 3.0 & -0.690 & 78.00 & 20.00 \\
128898 & 645 & 4.48 & 13 & 0.205 & 38.23 & 39.95 \\
133029 & 2420 & 2.11 & 25 & 0.417 & 26.70 & 39.28 \\
137909 & 673 & 18.49 & 2.0 & -0.600 & 15.00 & 85.00 \\
137949 & 1498 & 27450 & 4.0 & \multicolumn{1}{c}{--} & \multicolumn{1}{c}{--} & \multicolumn{1}{c}{--} \\
140160 & 230 & 1.60 & 64 & -0.366 & 51.95 & 59.34 \\
143473 & 4293 & 2.84 & 25 & 0.530 & 44.55 & 17.33 \\
148112 & 650 & 3.04 & 35 & 0.395 & 36.09 & 30.75 \\
148199 & 899 & 7.73 & 15 & -0.544 & 64.44 & 58.30 \\
148330 & 304 & 4.29 & 18 & -0.362 & 30.69 & 74.46 \\
152107 & 1487 & 3.87 & 35 & 0.262 & 88.56 & 0.84 \\
153882 & 1751 & 6.01 & 13.9 & -0.759 & 27.55 & 85.91 \\
164258 & 756 & 0.86 & 53 & -0.803 & 14.33 & 88.40 \\
170000 & 374 & 1.72 & 54 & -0.369 & 45.35 & 64.98 \\
170397 & 616 & 2.25 & 38 & -0.845 & 64.00 & 80.23 \\
187474 & 1488 & 2345 & 0 & -1.000 & 86.00 & 45.00 \\
188041 & 2226 & 223.8 & 4 & 0.380 & 70.00 & 10.00 \\
192678 & 1411 & 6.42 & 4.8 & 1.000 & 4.00 & 32.00 \\
196502 & 492 & 20.27 & 4.2 & -0.817 & 19.83 & 87.92 \\
201601 & 728 & 27027 & 0 & \multicolumn{1}{c}{--} & \multicolumn{1}{c}{--} & \multicolumn{1}{c}{--} \\
208217 & 1456 & 8.44 & 15.3 & -0.440 & 15.00 & 86.00 \\
220825 & 269 & 1.14 & 36 & -0.415 & 23.92 & 79.61 \\
\hline
\end{tabular}
\end{center}
\end{table}

\begin{figure}
\centering
{\includegraphics[width=0.45\textwidth]{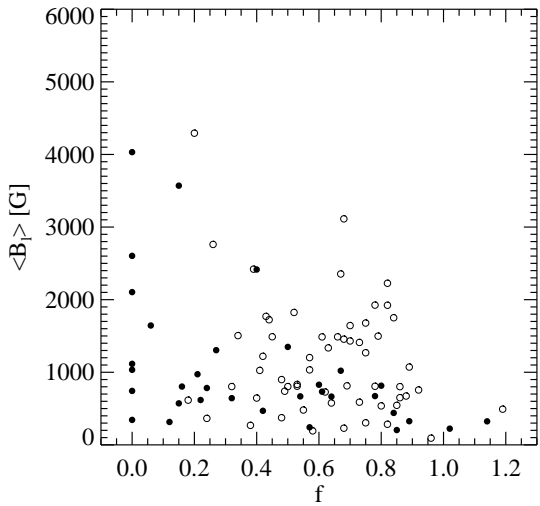}}
\caption{The rms longitudinal magnetic field as a function of the completed fraction of main-sequence life.
In this and the following figures, filled circles indicate stars with mass \highm{}, while open circles
indicate stars with mass \lowm{}.
}
\label{fig02}
\end{figure}

\begin{figure}
\centering
{\includegraphics[width=0.45\textwidth]{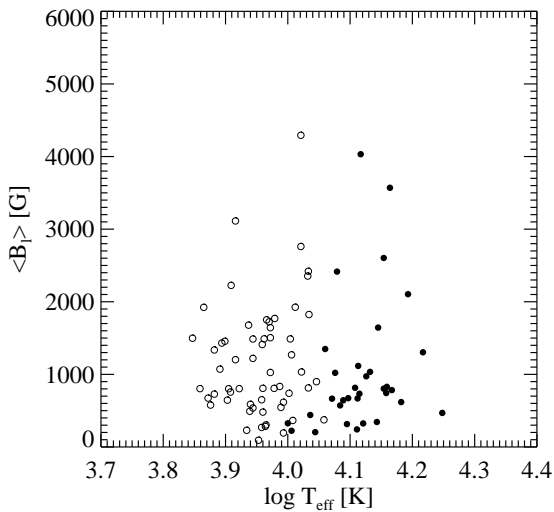}}
\caption{The rms longitudinal magnetic field as a function of effective temperature.
}
\label{fig03}
\end{figure}

\begin{figure}
\centering
{\includegraphics[width=0.45\textwidth]{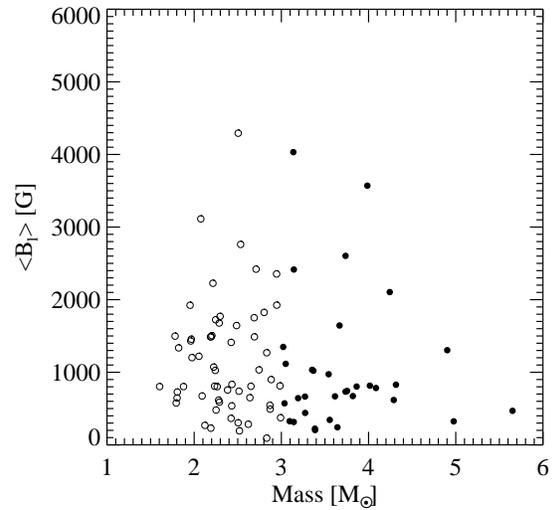}}
\caption{The rms longitudinal magnetic field as a function of mass.
}
\label{fig04}
\end{figure}

\begin{figure}
\centering
{\includegraphics[width=0.45\textwidth]{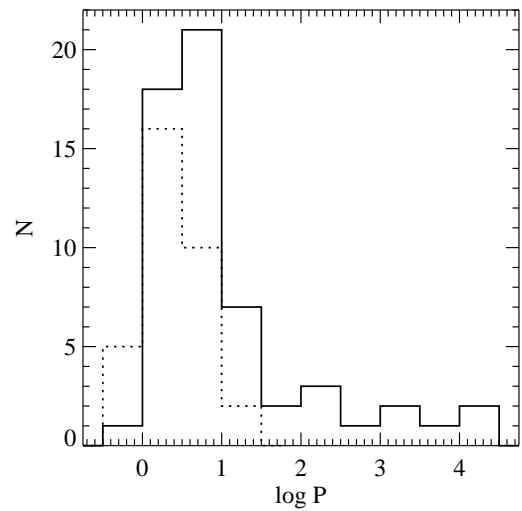}}
\caption{Distribution of $\log P$ for stars with masses \highm{} (dotted
line) and \lowm{} (solid line). 
}
\label{fig01b}
\end{figure}

\begin{figure}
\centering
{\includegraphics[width=0.45\textwidth]{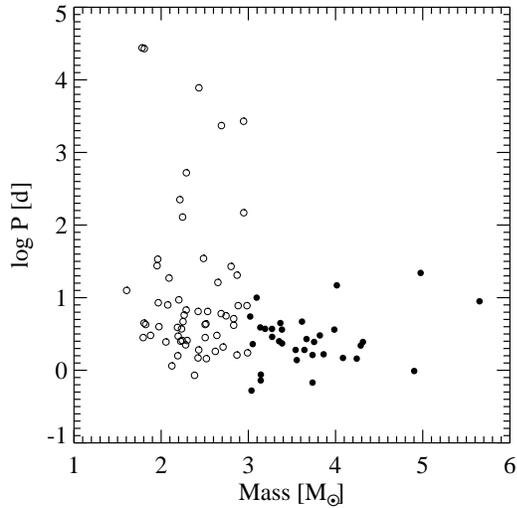}}
\caption{Rotation period as a function of mass.
}
\label{fig07}
\end{figure}


\begin{figure}
\centering
{\includegraphics[width=0.45\textwidth]{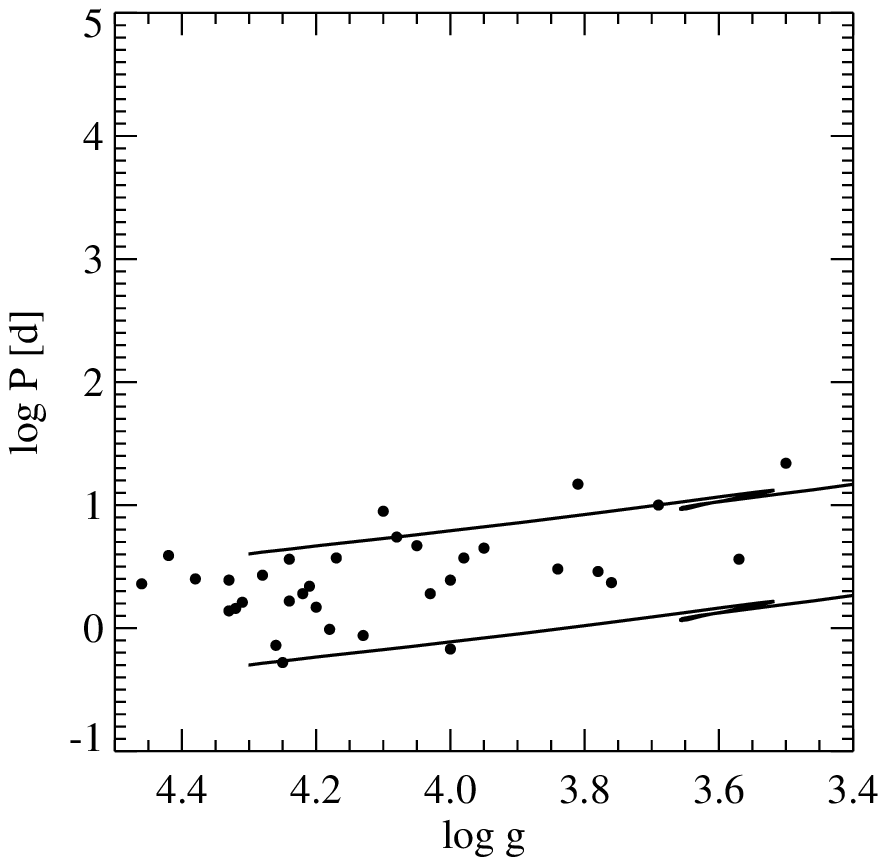}}
\caption{Rotation period as a function of surface gravity for stars with \highm{}.
The solid lines represent the theoretical evolution of the period for a 4\,\solarmass{} star and initial 
periods of 0.5 and 4~days, resulting from conservation of angular momentum alone.
}
\label{fig05a}
\end{figure}

\begin{figure}
\centering
{\includegraphics[width=0.45\textwidth]{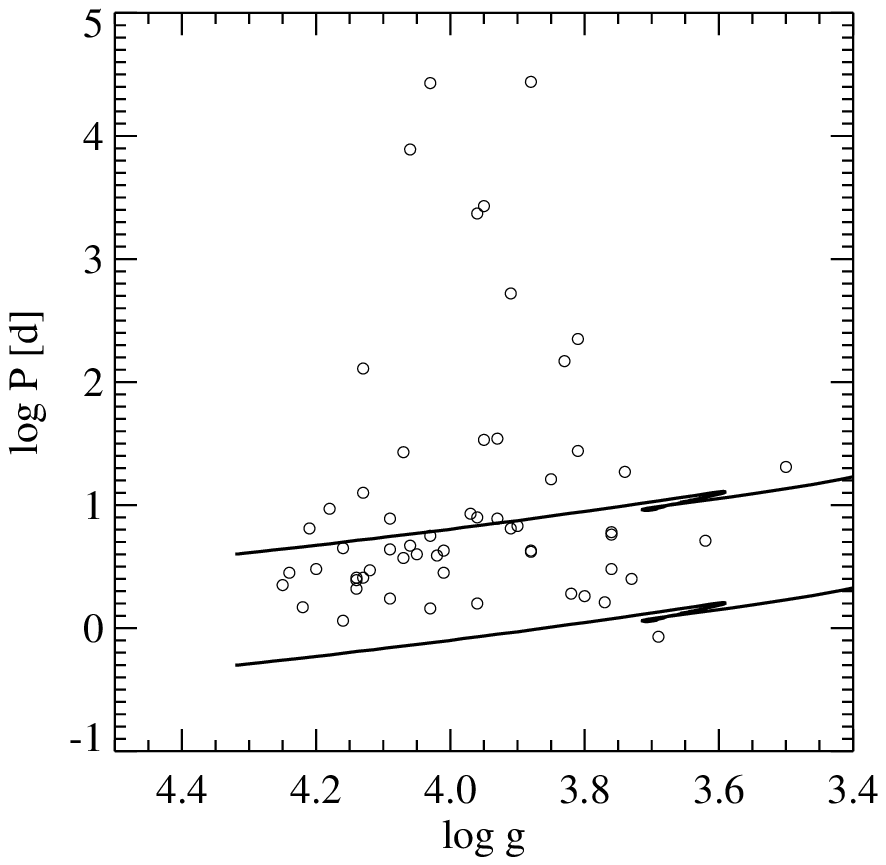}}
\caption{Rotation period as a function of surface gravity for stars with \lowm{}.
The solid lines represent the theoretical evolution of the period for a 2\,\solarmass{} star and initial
periods of 0.5 and 4~days, resulting from conservation of angular momentum alone.
}
\label{fig05b}
\end{figure}

\begin{figure}
\centering
{\includegraphics[width=0.45\textwidth]{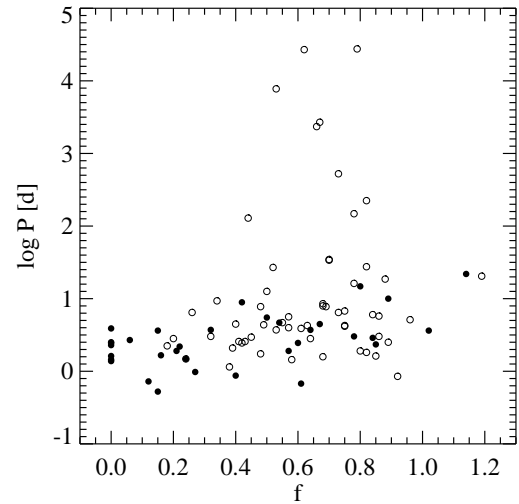}}
\caption{Rotation period versus elapsed time on the main sequence.
}
\label{fig14}
\end{figure}

\begin{figure}
\centering
{\includegraphics[width=0.45\textwidth]{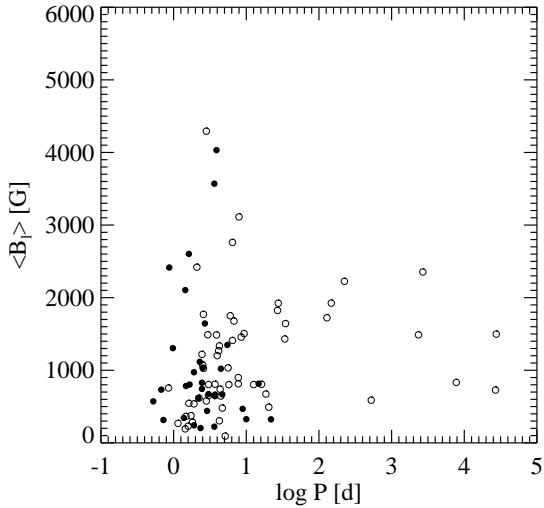}}
\caption{The rms longitudinal magnetic field as a function of period.
}
\label{fig44}
\end{figure}

\begin{figure}
\centering
{\includegraphics[width=0.45\textwidth]{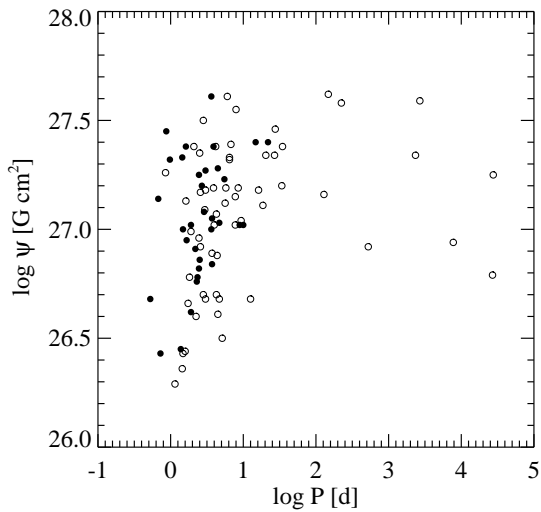}}
\caption{Magnetic flux against rotation period.
}
\label{fig10}
\end{figure}

\begin{figure}
\centering
{\includegraphics[width=0.45\textwidth]{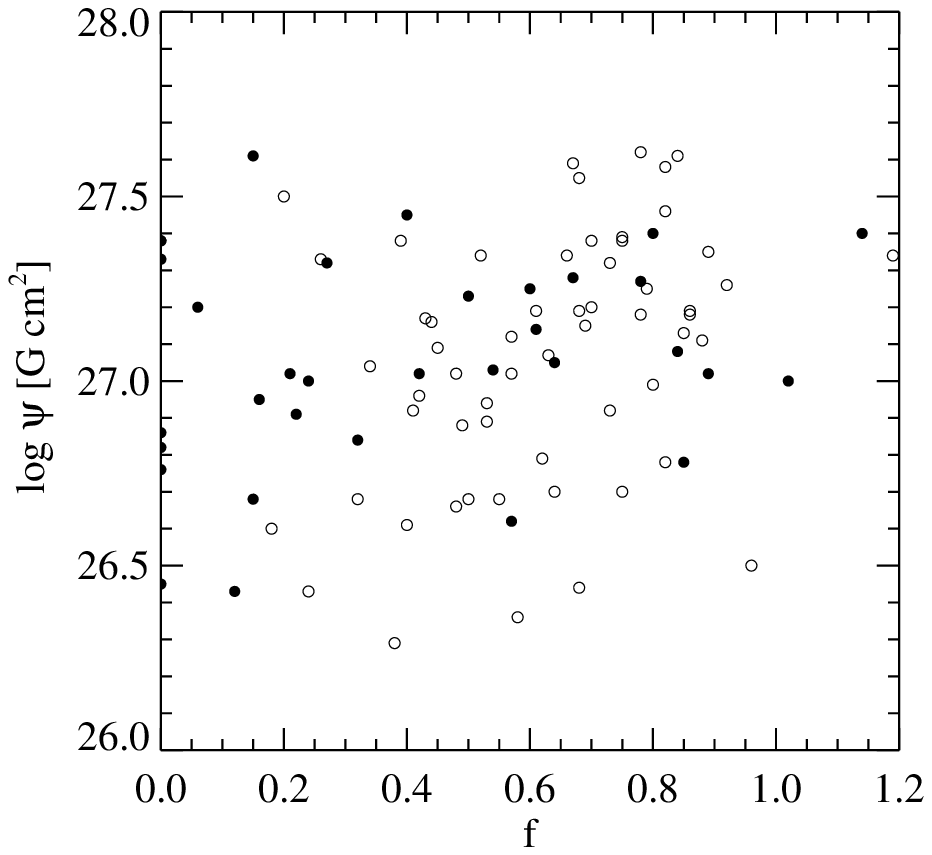}}
\caption{Magnetic flux against elapsed time on the main sequence.
}
\label{fig13}
\end{figure}

Because of the strong dependence of the longitudinal magnetic field on the rotational 
aspect, its usefulness to characterise actual magnetic field strength distributions is 
limited, but this is overcome by repeated observations to sample various rotation phases,
hence various aspects of the magnetic field. As an estimate of the magnetic field strength we used 
the rms longitudinal magnetic field strength compiled by Bychkov, Bychkova \& Madej (\cite{bych03}).
It has been computed from all measurements following the formula of Borra,
Landstreet \& Thompson (\cite{bo83}):

\begin{equation}
\overline{\left< B_l \right>} = \left( \frac{1}{n} \sum^{n}_{i=1} \left< B_l \right> ^2_i \right)^{1/2}.
\end{equation}

The $\overline{\left< B_l \right>}$ values are presented for each sub-sample separately in Tables~\ref{tab:3} and \ref{tab:4} 
in the second columns and have been preferentially selected from
magnetic field measurements carried out using Balmer hydrogen 
lines. As hydrogen is expected to be homogeneously distributed over the stellar surface, the 
longitudinal magnetic field measurements should sample the magnetic field fairly uniformly over 
the observed hemisphere.
In Fig.~\ref{fig02} the rms longitudinal magnetic field is plotted against the age of the stars 
expressed as a fraction of their total main-sequence life.
For the sub-sample of magnetic stars with \highm{} the strongest magnetic fields are found in younger stars 
in terms of the elapsed fraction of their main-sequence life. The fact that the strongest  magnetic fields 
are only observed close to the ZAMS can be interpreted as a magnetic field decay in 
stars at advanced ages. A similar decay of the magnetic field in stars with \highm{}
had been suggested more than twenty years ago by North \& Cramer (\cite{no84}) from the 
study of \logg{} values and surface magnetic fields estimated from Geneva photometry
(see their Fig.~6a). However, even though their result was based on hundreds of
stars, it had only a statistical and qualitative value, because the correlation
they used between a photometric parameter and the surface magnetic field is not a
one-to-one relation.
On the other hand, the stellar radii of stars increase by approximately a factor of two during 
the stellar life time across the main sequence. Therefore, rather than a real decay
of the magnetic field, our observations may simply testify for
conservation of the magnetic surface flux: in such a case,
the magnetic field strength is expected to have decreased by a factor of 4 by
the time the star reaches the TAMS. 

The magnetic field for the majority of the stars with \lowm{} becomes observable 
only after they have already completed a significant part of their life on the main sequence.
We note that the magnetic field strength obviously decreases after the low mass stars have 
finished 80\% of their main-sequence life. This would explain 
the rareness of detection of magnetic stars close to the TAMS.

In Figs.~\ref{fig03} and \ref{fig04} we present the rms longitudinal magnetic field as a function 
of effective temperature and mass, respectively. The distribution of stars in these diagrams reveals
that the strongest magnetic fields tend to be found in stars with an effective temperature between 
10,000\,K and 15,000\,K, and in the mass domain between 2.5 and 4\,\solarmass{}.

\subsection{Rotation}

As mentioned above, our sample includes exclusively Ap and Bp stars with well studied 
periodic magnetic field variations. The periods of variation are available in the catalog of Bychkov, Bychkova \& Madej
(\cite{bych05}) and present a compilation from the literature and in some cases their own determinations. 
The rotation periods P for each sub-sample are listed  separately in Tables~\ref{tab:3} and \ref{tab:4}
in the third column. Our previous study of the evolution of magnetic fields in Ap stars with well-known
surface magnetic fields (Hubrig, North \& Mathys 
\cite{hu00a}) suffered from a selection effect in the sense that the sample contained a 
high fraction (about 2/3) of stars with rotation periods longer than 10 days,
while the majority of the periods of magnetic stars fall between 2 and 4 days. 
For the present study the distribution 
of rotation periods is representative of that of all Ap and Bp stars.
In Fig.~\ref{fig01b} we show the distribution
of the periods for all stars in our sample. The majority of the studied stars have periods shorter than 4~days.
Interestingly, the longest rotation periods are found in stars in the mass domain between 
1.8 and 3\,\solarmass{} (Fig.~\ref{fig07})
and with \logg{} values  ranging from 3.80 to 4.13 (Figs.~\ref{fig05a} and \ref{fig05b}).
In Fig.~\ref{fig14} we show the rotation 
period as a function of the relative age of Ap and Bp stars. The longest periods are 
found only in stars which spent already more than 40\% of their main sequence life. 
For Bp and Ap stars with \highm{}, both Figs.~\ref{fig05a} and \ref{fig14} display
a slight increase of rotation period with 
age which is consistent with the assumption of conservation of angular momentum during their life 
on the main sequence, without any hint of a braking mechanism. A similar result had
already been obtained by North (\cite{no85}) from the study of smaller samples of Ap and Bp stars with 
spectroscopic and photometric estimations of \logg{}-values. North (\cite{nort98})
confirmed this trend in much the same way as here, except that his sample was not
selected on the basis of magnetic fields, but on the basis of a confirmed Bp or Ap
classification, photometric (or other) period and Hipparcos parallaxes.
The observed strength of magnetic fields in stars with various rotation periods 
is presented in Fig.~\ref{fig44}. This figure clearly shows that the strongest 
magnetic fields are found only in fast-rotating stars. All stars with periods exceeding 
10~days have $\overline{\left< B_l \right>}$ values less that 2400\,G.

In Figs.~\ref{fig10} and \ref{fig13} we present the magnetic flux $\Psi$
($= 12 \pi R^2 \overline{\left< B_l \right>}$,
assuming $B_s \approx 3 \overline{\left< B_l \right>}$),
plotted against the stellar rotation period and against the elapsed time on the main sequence.
Our Fig.~\ref{fig10} looks quite similar to Fig.~10 presented in our previous study of magnetic stars 
with well-determined surface magnetic fields (Hubrig, North \& Mathys \cite{hu00a}). 
 Although the stars with \highm{} show little
signs of correlation between magnetic flux and rotational period ---  although
one does have the visual
impression of a slight positive correlation --- the stars with \lowm{} do
show some trend. The flux for the latter stars seems to take random values 
for $\log P < 1.2$, while there is a slight decreasing trend
for $\log P > 1.5$. It is also remarkable that there is a lack of points in the quadrant 
$\log P > 1.2$ and $\log \Psi < 26.7$. The significance of this gap is attested
by a $2\times 2$ contingency table, using the limits just mentioned: the
two-sided Fischer exact probability value is $p=0.0251$, so we are comfortably
above the $95$\% significance level. On the other hand, an observational bias is
not completely excluded, because of the very long periods involved: it is
possible that the observers were motivated to follow only those stars with the
largest magnetic fields for such long periods of time (remember that we selected
here only stars with a decent magnetic curve). Thus the reality of the gap
remains to be proven without any doubt.
The plot of $\log \Psi$ against the elapsed 
fraction of the main sequence life shows significant dispersion and no sign of evolution of 
the magnetic flux.

\subsection{Cluster members with accurate Hipparcos parallaxes}

\begin{table*}
\caption{
Fundamental parameters for cluster stars with accurate Hipparcos parallaxes.
}
\label{tab:5}
\begin{center}
\begin{tabular}{rrrrrrrrrr}
\hline
\multicolumn{1}{c}{HD} & \multicolumn{1}{c}{M$_V$} & \multicolumn{1}{c}{$M$/M$_\odot$} & \multicolumn{1}{c}{log $T_{\rm eff}$} & \multicolumn{1}{c}{log $L$/L$_\odot$} & \multicolumn{1}{c}{\logg} & \multicolumn{1}{c}{$R$/R$_\odot$} & \multicolumn{1}{c}{d[pc]} & \multicolumn{1}{c}{$\sigma(\pi)/\pi$} & \multicolumn{1}{c}{$f$} \\
\hline
23408 & -1.93 & 4.463$\pm$0.104 & 4.079$\pm$0.013 & 2.916$\pm$0.053 & 3.44$\pm$0.07 & 6.66$\pm$0.57 & 132. & 0.040 & 1.11 \\
23950 & 0.07 & 3.401$\pm$0.103 & 4.112$\pm$0.013 & 2.167$\pm$0.053 & 4.20$\pm$0.07 & 2.42$\pm$0.21 & 132. & 0.040 & 0.26 \\
63401 & -0.37 & 3.480$\pm$0.170 & 4.129$\pm$0.013 & 2.168$\pm$0.103 & 4.28$\pm$0.10 & 2.24$\pm$0.30 & 219. & 0.109 & 0.07 \\
74196 & -0.40 & 3.604$\pm$0.114 & 4.108$\pm$0.013 & 2.352$\pm$0.056 & 4.03$\pm$0.08 & 3.05$\pm$0.27 & 153. & 0.046 & 0.57 \\
92385 & 0.75 & 2.730$\pm$0.106 & 4.043$\pm$0.013 & 1.804$\pm$0.077 & 4.19$\pm$0.09 & 2.19$\pm$0.23 & 151. & 0.076 & 0.30 \\
92664 & -0.32 & 3.863$\pm$0.147 & 4.154$\pm$0.013 & 2.357$\pm$0.075 & 4.24$\pm$0.09 & 2.47$\pm$0.26 & 143. & 0.073 & 0.16 \\
108945 & 0.57 & 2.431$\pm$0.085 & 3.944$\pm$0.015 & 1.725$\pm$0.079 & 3.82$\pm$0.09 & 3.16$\pm$0.36 & 95. & 0.079 & 0.80 \\
\hline
\end{tabular}
\end{center}
\end{table*}

Another way to get an insight into the evolution of stellar magnetic fields is to study magnetic 
stars in open clusters or associations at different ages.
As far as the membership of Ap stars in 
distant open clusters or associations is concerned,
we should keep in mind that such studies are mostly based upon photometry
and upon radial velocity determinations. However, criteria for assessing 
cluster membership based on photometry cannot be applied to peculiar stars
straight away, since
strong backwarming effects lead to an anomalous energy distribution, thus
affecting the position of the stars in colour--magnitude diagrams.

In our sample, seven Ap/Bp stars are known members of nearby open clusters of different ages and 
have very accurate Hipparcos parallaxes (see Table~\ref{tab:5}).
However, among them, only two stars, HD\,92664 and HD\,108945 
have well-defined magnetic rotational phase curves (Bychkov, Bychkova \& Madej \cite{bych05}).
The magnetic field 
measurements with FORS\,1 at the VLT allowed to detect a magnetic field of the 
order of $-$600\,G in the spectra of two other stars, HD\,63401 and HD\,92385 (Hubrig et al.\ \cite{hu06}).
On the other hand, we failed to measure magnetic fields in the remaining three 
stars, HD\,23408, HD\,23950 and HD\,74196. It is quite possible that they are not typical 
magnetic Ap/Bp stars, as they show in their spectra an overabundance of Mn and Hg and strong deficit  of He.
In summary, we are left with four cluster members which seem to represent 
typical magnetic stars.

For most cluster stars, we preferred to use the distance given by main
sequence fitting over the Hipparcos parallax. The error is generally assumed
to be $\pm$0.10 magnitudes on the distance modulus. For the Pleiades, of which 
HD\,23408 and HD\,23950 are members, the adopted
distance was $132\pm4$\,pc, according to the geometric estimate by
Zwahlen et al.\ (\cite{zwa04}) for the Atlas binary,
in agreement with the estimate of Munari et al.\ (\cite{mun04}) for the eclipsing binary HD\,23642 ($132\pm2$\,pc).
For NGC 2451, of which HD\,63401 is a member,
we have adopted the distance modulus of Maitzen \& Catalano (\cite{maca86}): (m-M)$_0=6.7$ and no reddening. 
For IC 2391, of which HD\,74196 is a member, and for IC 2602, the true
distance moduli were adopted from North (\cite{nor93}). 
For HD\,108945 (in Coma Berenices), we just used the Hipparcos parallax, because of 
a possible depth effect.

\begin{figure}
\centering
{\includegraphics[width=0.45\textwidth]{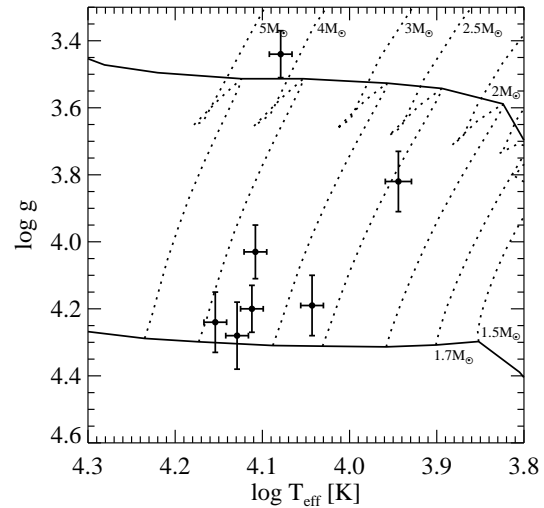}}
\caption{H-R digram for cluster stars. The vertical and horizontal 
bars indicate the accuracy of the determination of effective temperature and gravity.}
\label{fig00cl}
\end{figure}

The fundamental parameters for the cluster stars are presented in Table~\ref{tab:5}, and their 
position in the H-R diagram is shown in Fig.~\ref{fig00cl}.
Both HD\,92385 and HD\,108945 with \lowm{} spent already 
30 and 80\% of their main sequence life, respectively, whereas HD\,63401 and HD\,92664 with
\highm{} are very young and with a completed fraction of their main sequence life of
respectively 7 and 16\%.
These results are in full agreement with conclusions made in the previous section in 
relation to the age of low and high mass stars.

Only a few double--lined spectroscopic binary systems  (which can be considered as very small 
clusters) containing a magnetic Ap star are currently known,  and in all studied systems
the Ap components with \lowm{} have already completed a significant fraction of their 
main-sequence life (e.g., Wade et al.\ \cite{wa96}, North et al.\ \cite{no98},
Carrier et al.\ \cite{ca02}).

\subsection{The magnetic field geometry}

There are two different ways to infer the geometry of the magnetic fields. Obviously, the 
observed magnetic curves can be modeled by assuming that the magnetic geometry can be 
described by the superposition of a dipole and a quadrupole magnetic field or higher multipoles with 
arbitrary orientation.
However, we should keep in mind that generally different magnetic field distributions and 
different inclination angles $i$ between the rotation axis and the line of sight and  
inclination angles $\beta$ between the magnetic axis and the rotation axis can represent 
the observed magnetic curves equally well. 
On the other hand, the vast majority of the magnetic CP stars exhibits a smooth, single-wave variation of 
the longitudinal magnetic field during the stellar rotation cycle. These observations are considered as evidence 
for a dominant dipolar contribution to the magnetic field topology.
A first statistical investigation of the orientation of 
magnetic axes in periodic magnetic variables based on the assumption that the studied 
stars are oblique dipole rotators has been carried out by Preston (\cite{pre67})
with the result that their magnetic and rotation axes are nearly orthogonal.
He defined

\begin{equation}
r = \frac{\left< B_l\right> (min)}{\left< B_l\right> (max)}
  = \frac{\cos \beta \cos i - \sin \beta \sin i}{\cos \beta \cos i + \sin \beta \sin i},
\end{equation}

\noindent
so that

\begin{equation}
\beta =  \arctan \left[ \left( \frac{1-r}{1+r} \right) \cot i \right].
\label{eqn:4}
\end{equation}

Observations of spectral 
line widths and profiles allow to determine the projected rotation velocity $v$\,sin\,$i$, 
where $v$ is the equatorial rotation velocity and $i$ is the inclination angle of the 
rotation axis. The equatorial rotation velocity is given by $v_{\rm eq} = 50.6 R/P$, where $R$ is the 
stellar radius in solar units and $P$ the period in days. From the measured $v$\,sin\,$i$ 
values and $v_{\rm eq}$ determined using the known rotation period and the radius computed from the 
luminosity and effective temperature, the inclinations of the stellar rotation axes can 
easily be calculated. 
Knowing $i$ and $r$, the determination of the obliquity angle $\beta$ for the orientation of the 
magnetic axis with respect to the rotation axis is rather straightforward.

For our study, the quantities $r$ for all sample stars, except for HD\,81009 have been taken from 
the catalogue of stellar magnetic rotational phase curves compiled by
Bychkov, Bychkova \& Madej (\cite{bych05}).
For HD\,81009 we used $r$ = 0.64, determined by  Landstreet \& Mathys (\cite{lama00}).
The $v$\,sin\,$i$ values have been measured on our spectra or taken from
published spectroscopic studies 
(Abt \cite{abt01},
Abt, Levato \& Grosso \cite{abt02},
Bagnulo et al.\ \cite{bag02},
Balona \& Laney \cite{bal03},
Bohlender, Landstreet \& Thompson \cite{boh93},
Hensberge et al.\ \cite{hen79},
Landstreet \cite{lan70},
Landstreet \& Mathys \cite{lama00},
Levato et al.\ \cite{lev96},
Nielsen \& Wahlgren \cite{nie02},
Royer et al.\ \cite{roy02},
Wade et al.\ \cite{wad00}).
We could not find in the literature values of $v$\,sin\,$i$ for three stars: HD\,54118, HD\,73340 and 
HD\,133880.
The inclination 
angles $i$ between the rotation axis and the line of sight determined from $v$\,sin\,$i$ and $v_{\rm eq}$
have been used to calculate the obliquities $\beta$ following equation~\ref{eqn:4}.

In columns 4 to 7 of tables~\ref{tab:3} and \ref{tab:4}, we present for each star in our sub-samples 
respectively the values 
$v$\,sin\,$i$, the parameter $r$ from the catalog of Bychkov, Bychkova \& Madej \cite{bych05} and the 
inclinations $i$ and $\beta$.
For magnetic Ap stars with rotation periods of the order of one month or longer, the 
computed  $v_{\rm eq}$ are very small
and the inclination angle $i$ between the rotation 
axis and the line of sight cannot be determined. A couple of years ago Landstreet \& 
Mathys (\cite{lama00}) carried out modeling of observed magnetic field
moments for a small sample of slowly 
rotating magnetic stars. Their result was rather unexpected, 
implying that the magnetic stars with periods P$>$25\, days have small values of 
obliquity $\beta$ of the model magnetic axis to the rotation axis, of the order of 
20$^\circ$. A few stars from their study, which have accurate Hipparcos parallaxes and 
available Str\"omgren and/or Geneva photometry have been included in our sample.
In fact, we simply employed the magnetic field parameters determined by these authors 
in our study to increase the statistical significance of our results.

\begin{figure}
\centering
{\includegraphics[width=0.41\textwidth]{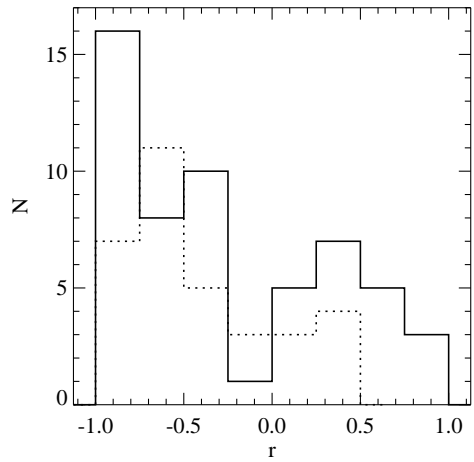}}
{\includegraphics[width=0.41\textwidth]{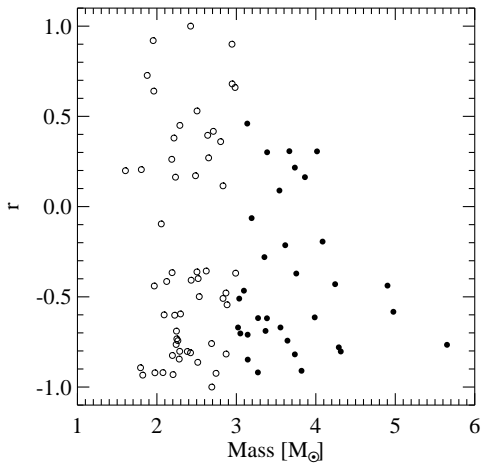}}
{\includegraphics[width=0.41\textwidth]{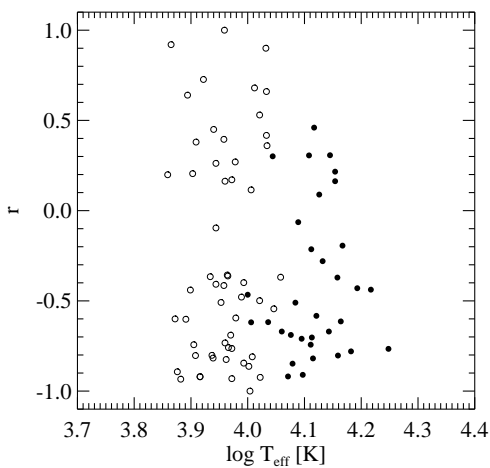}}
\caption{Upper panel: Distribution of $r$-value for stars with masses \highm{} (dotted
line) and \lowm{} (solid line). Middle panel: $r$ value versus mass. Lower panel: $r$-value
versus effective temperature.
}
\label{fig06}
\end{figure}

\begin{figure}
\centering
{\includegraphics[width=0.45\textwidth]{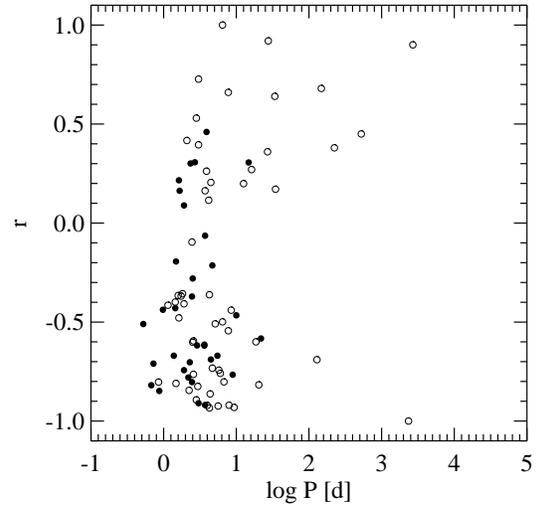}}
\caption{$r$-value versus rotation period.
}
\label{fig09}
\end{figure}


\begin{figure}
\centering
{\includegraphics[width=0.45\textwidth]{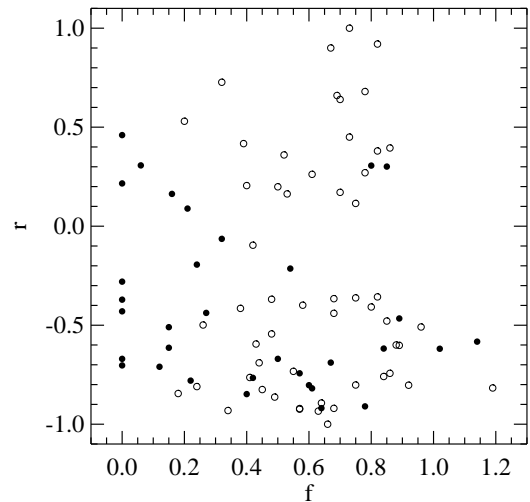}}
\caption{The $r$-value versus elapsed time on the main sequence.
}
\label{fig22}
\end{figure}

\begin{figure}
\centering
{\includegraphics[width=0.45\textwidth]{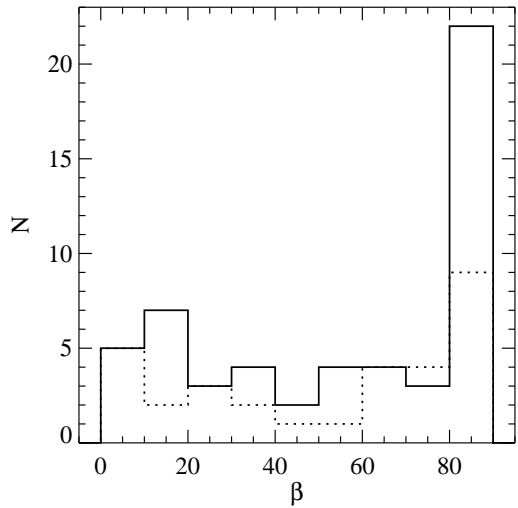}}
\caption{Distribution of obliquity angles $\beta$ for stars with masses \highm{} (dotted
line) and \lowm{} (solid line). 
}
\label{fig01a}
\end{figure}


\begin{figure}
\centering
{\includegraphics[width=0.45\textwidth]{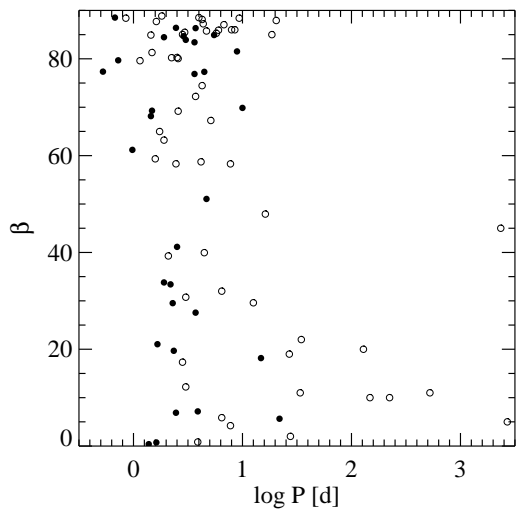}}
\caption{$\beta$ versus rotation period.
}
\label{fig11}
\end{figure}


\begin{figure}
\centering
{\includegraphics[width=0.45\textwidth]{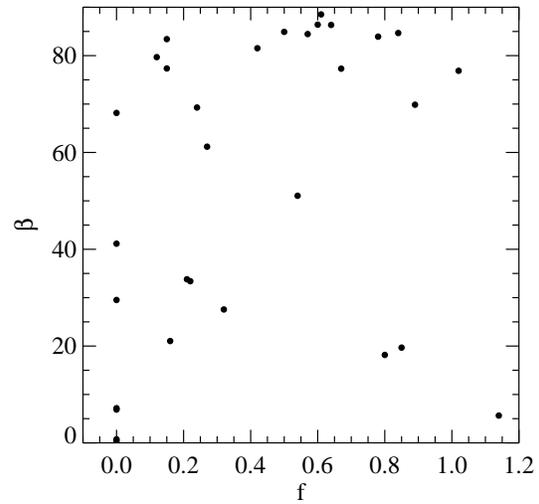}}
{\includegraphics[width=0.45\textwidth]{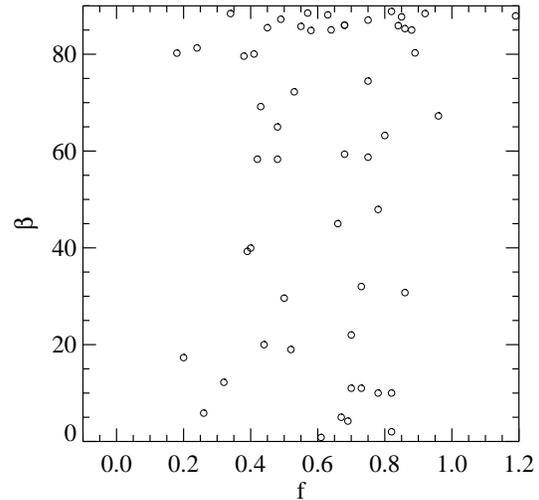}}
\caption{$\beta$-value for high (upper panel) and low (lower panel) mass stars versus 
elapsed time on the main sequence.
}
\label{fig15} 
\end{figure}


\begin{figure}
\centering
{\includegraphics[width=0.45\textwidth]{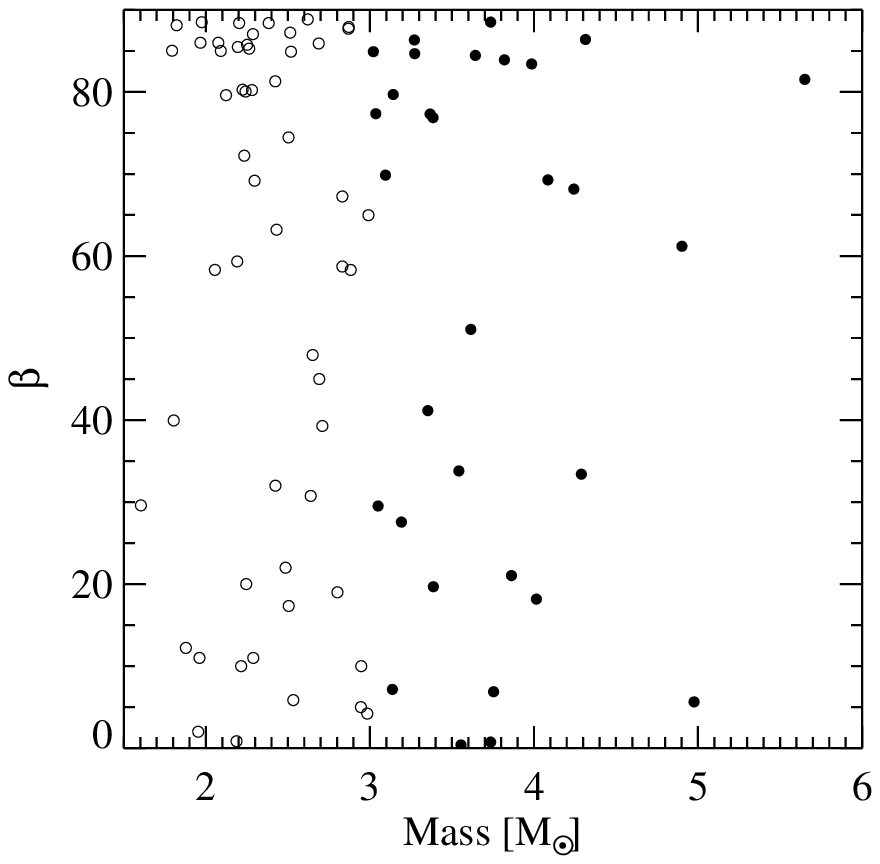}}
{\includegraphics[width=0.45\textwidth]{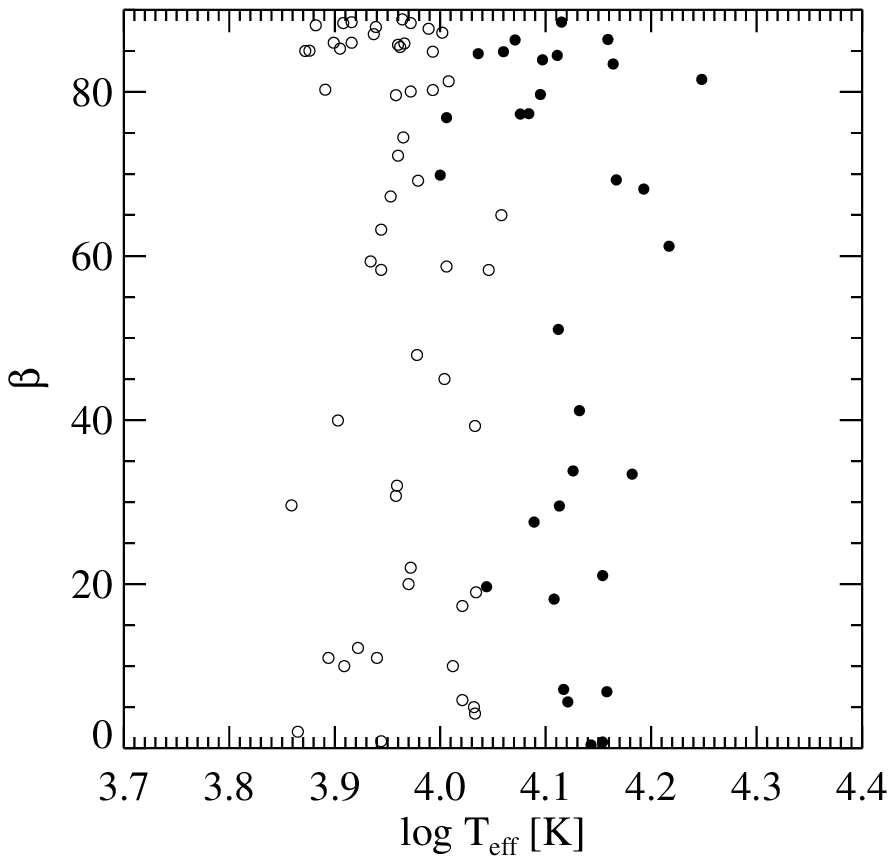}}
\caption{
$\beta$-value versus mass (upper panel) and $\beta$-value versus effective temperature (lower panel).
}
\label{fig17}
\end{figure}

In the upper panel of Fig.~\ref{fig06} we present the distribution of $r$ values for stars of
\highm{} and of \lowm{} shown by a dotted and a solid line, respectively.
The distributions look rather different with an approximately bimodal distribution 
for lower mass stars with $r$ peaked between $-$0.75 and $-$1 (highest peak) and between 0.25 and 0.5. The 
highest peak in the distribution of $r$ values for stars with \highm{} is shifted to higher values
between $-$0.75 and $-$0.5. Both distributions nearly resemble the $r$ value distribution 
presented in older studies of Preston (\cite{pre67,pre71}) and 
Landstreet (\cite{lan70}), with an apparent deficit of stars with intermediate values of $\beta$. 
According to probability distributions of $r$ calculated 
by Landstreet (\cite{lan70}), we conclude that the majority of stars in our sample must have a $\beta$ larger 60$^\circ$.
The lack of stars with $r$-values near zero can be explained either by an observational effect or 
by a lack of small $\beta$-values. 
Further, we note that no star with \highm{} has $r$ larger than 0.5, indicating the difficulty 
of detecting periodicity in stars with very small magnetic field variations.
The examination of the middle panel of Fig.~\ref{fig06} with 
$r$ values plotted against stellar mass reveals no obvious dependence for lower mass stars whereas 
higher mass stars with \highmx{4}{} show a slight trend to have $r<-0.5$. This behaviour can 
be interpreted by the prevalence of larger obliquities $\beta$ in more massive stars.
From the inspection of the dependence between the $r$-values and effective temperatures 
(lower panel in Fig.~\ref{fig06}) the same is possibly true for the hottest stars in our sample, i.e.
larger obliquities $\beta$ are found in hotter stars.
 
Quite interesting results follow from the study of 
the distribution of $r$-values for stars with different rotation periods 
and the distribution of $r$-values over the age, i.e.\ elapsed time on the main 
sequence presented in Figs.~\ref{fig09} and \ref{fig22}.
While we observe a random distribution of obliquity angles $\beta$ in faster 
rotating stars, it seems that long period magnetic variables of stars with \lowm{} tend to have 
values of $\beta$ close to 0$^\circ$. 
In the distribution of $r$-values over the age we observe that the obliquity angle distribution 
as inferred from the distribution of $r$-values
appears random at the time magnetic stars become observable on the H-R diagram. After quite 
a short time spent by a star with mass \lowm{} on the  main sequence the 
obliquity angle $\beta$ tends to reach  values 
close to either to 90$^\circ$ or 0$^\circ$. In other words, the magnetic axis becomes either 
perpendicular to the rotation axis or aligned with it with advanced age. 
A similar result revealing the bimodal distribution based on a smaller 
sample of Ap and Bp stars had been presented by North (\cite{no85}), but on the
basis of less reliable photometric \logg{} estimates.
For more massive stars with \highm{} the obliquity angle $\beta$  tends to reach  values 
close to 90$^\circ$.

The distribution of obliquity angles $\beta$  for stars with \highm{} and 
\lowm{} is presented in Fig.~\ref{fig01a}. In agreement with 
the discussion on the distribution of $r$-values, the excess of stars with large $\beta$-values is 
detected in both higher and lower mass stars.
Recently, Bychkov, Bychkova \& Madej (\cite{bych04}) presented a study of parameters of magnetic phase curves 
based on their own compilation in Bychkov, Bychkova \& Madej (\cite{bych05}) of 
the data for 134 Ap and Bp stars. Surprisingly, they found an excess of stars with 
angles $\beta$ close to 0$^\circ$. We do not know the origin of this 
discrepancy between their and our study, but we only emphasise here that our sample consists 
of carefully selected 90 Ap and Bp stars with a sufficient number of magnetic field 
measurements and well-defined periods and magnetic phase curves. 
From the inspection of 
plots displaying the distribution of $\beta$ in fast and slowly rotating stars (Fig.~\ref{fig11}),
it is obvious that large $\beta$ ($\ge$80$^\circ$) outweigh the distributions of 
faster rotating stars whereas the
angle $\beta$ is smaller than 20$^\circ$ in slower rotating stars. 
We note that all nine points with small values of $\beta$ in slowly rotating stars are from the study 
of Landstreet \& Mathys (\cite{lama00}). As for stars with \highm{}, the two stars with the longest 
periods (HD\,5737 and HD\,168733) have magnetic and rotation axes aligned to within 20$^\circ$, 
similar to the behaviour of slowly rotating stars of lower mass. Clearly, more 
magnetic field studies of slowly rotating stars with higher mass are needed to confirm this behaviour.

The evolution of the obliquity angle $\beta$ seems to be quite different for 
low and high mass stars (Fig.~\ref{fig15}). We find a strong hint 
for an increase of $\beta$ with elapsed time on the main sequence for stars 
with \highm{}. However, no similar trend is found for lower mass stars, although 
the predominance of high values of $\beta$ at advanced ages is notable. 
Further, we do not find any trend for $\beta$-values either with mass or with \teff{}
(Fig.~\ref{fig17}).

As is mentioned by Bychkov, Bychkova \& Madej (\cite{bych05}), a number of stars in their catalog
exhibit marked anharmonicity, which indicates significant departures from dipole geometry.
To reproduce the shapes of the observed longitudinal magnetic field variation, the magnetic field 
distribution has to be regarded as a sum of dipole and quadrupole components. 
Out of the 33 stars studied with \highm{}, 21\% have magnetic phase curves
fitted by a double wave,
whereas the percentage of stars with \lowm{} for which the magnetic field variations 
had to be fitted by a double wave is 12\%.
This statistics could indicate the possibility that the magnetic topology in stars with \highm{}
is frequently more complex than that in less massive stars. The double-wave longitudinal field curves 
in stars with \lowm{} are mostly found at the age of 40-50\% of their main sequence life, while
a number of stars with \highm{} exhibit a non-dipolar field already at a very young age close to the ZAMS.
However, further systematic studies of mean longitudinal magnetic fields should be conducted to better assert 
the evolution of the magnetic field distributions
during the main-sequence life. Moreover, the observations of other magnetic field moments such as the mean quadratic 
field and mean crossover field, along with studies of Zeeman structure in the individual line profiles 
in circular and linear polarisation should be involved in future modeling to obtain a most 
accurate determination of the magnetic field topologies.

\section{Discussion}

We believe that the work presented here has
important implications for the understanding of the origin of the magnetism detected in 
A and B stars.
The new results based on the statistical properties of 90~Ap and Bp 
stars can be summarised as follows:

\begin{itemize}
\item The present study of the evolutionary state of upper main sequence magnetic stars
indicates a notable difference between the distributions of high mass and low mass magnetic 
stars in the H-R diagram.
In contrast to magnetic stars of mass
\lowm{} which are mostly  found around the centre of the main-sequence band, 
the stars with masses \highm{} seem to be concentrated closer to the ZAMS, and the stronger 
magnetic fields tend to be found in hotter, younger (in terms of the elapsed fraction of main-sequence
life) and more massive stars.

\item The longest rotation periods are found in stars in the mass domain between 
1.8 and 3\,\solarmass{}
and with \logg{} values  ranging from 3.80 to 4.13.
The longest periods are 
found only in stars which spent already more than 40\% of their main sequence life. 

\item No evidence is found for any loss of angular momentum during the main-sequence life
of stars with \highm{}.

\item No sign of evolution of the magnetic flux
over the stellar life time on the main sequence
has been found, in agreement with the assumption of magnetic flux conservation. 

\item Our study of a few known members of nearby open clusters of different ages and 
with accurate Hipparcos parallaxes confirms the conclusions based on the study of field stars 
with accurate Hipparcos parallaxes.

\item The excess of stars with large obliquities $\beta$ is detected in both higher and lower mass stars.
It is quite possible that the angle $\beta$ becomes close to 0$^\circ$ in slower rotating stars of 
high mass too, analog to the behaviour of angles $\beta$ in slowly rotating stars of \lowm{},

\item The obliquity angle distribution as inferred from the distribution of $r$-values
appears random at the time magnetic stars become observable on the H-R diagram. After quite 
a short time spent by a star on the  main sequence the obliquity angle $\beta$ tends to reach  values 
close to either 90$^\circ$ or 0$^\circ$ for \lowm{}. However, for \highm{} the obliquity 
angle $\beta$ in increasing with the completed fraction of the main-sequence life. 
It would be important to consider such a  behaviour 
in the framework of dynamo or fossil magnetic theories,
taking into account the influence of 
various mechanisms (e.g.\ possible changes in internal rotation during the evolutionary stage).

\item
While we find a strong hint 
for an increase of $\beta$ with the elapsed time on the main sequence for stars 
with \highm{}, the trend for lower mass stars is much less pronounced, though it
seems to exist as well.
\end{itemize}

The presented results have to be implemented in the long-lasting discussions about fossil or 
dynamo origin for the observed magnetic fields. 
It is quite possible that the observed magnetic fields in stars with \highm{},
showing the prevalence of 
stronger magnetic fields in younger stars, are in some sense fossil, being the remnants 
of magnetic fields originally present in the material from which these stars formed.
Recent theoretical work indicates that dynamo action is possible, even likely, in a convective 
core, but the overlying stellar radiative envelope represents a significant 
impediment to the appearance of any of the generated magnetic field at the surface 
(e.g., MacGregor \& Cassinelli \cite{mac03}, MacDonald \& Mullan \cite{mul04}).
To remedy the difficulties associated with the transport of magnetic fields from the core to the 
surface there have been a number of attempts to identify mechanisms that are capable of 
generating a magnetic field within the radiative envelope itself
(e.g., Spruit \cite{spr02}, Braithwaite \cite{bra04}, Arlt, Hollerbach \& R\"udiger \cite{arl03}). 
For stars with mass \lowm{} the magnetic fields appear after they completed
more than 20-30\% of their main-sequence life. Because of this apparent difference in the 
evolutionary state between higher mass and older lower mass stars it is tempting to 
make a conclusion that magnetic fields in stars of lower mass are generated by dynamo action.
However, the comparison of the observed magnetic field evolution and magnetic field geometries in 
both samples presented in this paper does not reveal any striking differences apart from the 
older age of lower mass magnetic stars on the 
main sequence, the very fact that most extreme slow rotators are found among stars with 
\lowm{}, and the difference in the evolutionary behaviour of obliquities between both samples.

\acknowledgements
We thank M.\ Netopil and E.\ Paunzen for useful discussions on membership of 
our sample stars in nearby open clusters.

\newpage

%

\end{document}